\documentclass[english,floatfix]{elsarticle}
\usepackage[LGR,T1]{fontenc}
\usepackage[utf8]{inputenc}
\usepackage{array}
\usepackage{amsmath}
\usepackage{amssymb}
\usepackage{graphicx}

\makeatletter

\providecommand{\tabularnewline}{\\}

\usepackage{float}

\ProvideTextCommand{\~}{LGR}[1]{\char126#1}

\usepackage{babel}

\makeatother

\usepackage{babel}
\begin{document}

\begin{frontmatter}{}

\title{$f(R)$ gravity with broken Weyl gauge symmetry, cosmological backreaction,
and its effects on CMB anisotropy}

\author{Jiwon Park}

\ead{cosmosapjw@soongsil.ac.kr}

\author{Tae Hoon Lee}

\address{Department of Physics, Soongsil University, Seoul 06978, Korea}
\begin{abstract}
We propose a new class of $f(R)$ theory where its Weyl gauge symmetry
is broken in the primordial era of the universe. This symmetry
forces one to adopt a new scalar field, namely a Weyl field and a gauge vector boson. Furthermore, an equivalent
form of the Einstein-Hilbert Lagrangian with a non-minimally coupled
scalar field corresponding to the function $f(R)$ is found. Due to
the geometrical feature of the Weyl field, it turns out that the symmetry
breaking induces a non-minimal coupling, which cannot be expected
in the standard $f(R)$ theories. We explain how this affects the evolution
of the universe at cosmological scales. It is shown that there may
be a value shift in the Planck constant and the cosmological constant.
This can be regarded as a genuine exemplification of the cosmological
backreaction. Furthermore, one also finds new features in the evolution
of perturbational variables and cosmic microwave background anisotropy.
Moreover, we prove that when a specific $f(R)$ model invokes 
inflation, the amplitude of the primordial gravitational waves affects
the evolution of scalar perturbation due to the new
non-minimal coupling. As a case study, we explain how this can be
embodied in the Starobinsky inflation. Finally, we discuss some impacts
that this  physics can bear and the possibility of giving
a new restriction of the estimation of cosmological variables such
as the gravitational wave amplitude with experiments.
\end{abstract}

\end{frontmatter}{}

\section{Introduction}

Symmetries and their breaking phenomena are essential concepts in
modern physics. Especially after the theoretical prediction \citep{1}
and experimental discovery \citep{2} of the Higgs particle, research
on symmetry breaking has become one of the main study areas of particle
physics. The idea of symmetry breaking has been vigorously studied
also in cosmology \citep{3,4,5,6,7,8,9}. However, there have not
been enough attempts to apply the concept to the geometrical symmetry
of spacetime itself in the cosmological context. For example, \citep{10}
studies how the diffeomorphism may be broken explicitly or spontaneously
in some modified gravity models like Chern-Simons gravity and bimetric
gravity and investigate their post-Newtonian limits but did not consider
the cosmological situation. The diffeomorphism invariance in $f(R)$
gravity was studied in \citep{11}, but the idea of symmetry breaking
was not applied. Hence, we are motivated to explore how symmetry breaking
may affect modified cosmological models, especially in the primordial
era.

Meanwhile, a new class of theories satisfying a new geometrical symmetry
known as Weyl geometry has attracted the interest of many cosmologists.
It originated from the philosophy that the physical theory should
allow and include the most considerable gauge symmetry possible
\citep{12,13}. In this context, Weyl geometry can be viewed as an
extension of Einstein's general relativity(GR) with just a bigger
gauge group \citep{14}. On the other hand, it can be derived from
the Brans-Dicke theory when one applies the Palatini formalism, which
claims that the metric and the connection are independent of each
other \citep{15,1703_05649}. Since the idea has many relationships with various
research themes of cosmology, many studies have applied it to many
problems of cosmology, such as dark energy and inflationary cosmology \citep{1703_05649,16,17,18,19,20}.

Recently, We attempted to apply the idea of symmetry breaking in Weyl
geometry. In \citep{21}, we have shown the possibility of distinguishing
two types of primordial symmetry breaking, model A and B, with an
observation of cosmic microwave background(CMB) anisotropy observables.
Model A originates from Zee's broken-symmetric theory of gravity \citep{22},
and model B applies Palatini formalism to model A. Significantly,
the symmetry breaking of Weyl geometry appeared in model B, which has a
particular interest because it has a new salient geometrical feature
that the standard explanation for symmetry breaking cannot have.

In this study, we expand this previous study on geometrical symmetry
breaking in Weyl geometry to the case where the Einstein-Hibert action
is replaced with an arbitrary scalar curvature function, i.e., $f(R)$
gravity. Currently, many $f(R)$ gravity models are vigorously studied
in cosmology and quantum gravity. These include attempts to find an appropriate
inflation model and search for a possible resolution for the dark
energy problem \citep{23,Nojiri2011,S.Nojiri2017}. In this context,
it would be worth studying how the geometrical symmetry breaking
has a distinctive effect on these subjects. It is especially shown that there can be a shift in values of fundamental physical constants such as the Planck constant. Furthermore, we explore
what consequences in the evolution of the universe and CMB observables this modified $f(R)$ gravity
leads by studying the Integrated-Sachs-Wolfe(ISW) effect of CMB temperature anisotropy spectra. For $f(R)$ inflation models, it is found that the amplitude
scale of primordial gravitational waves(GW) brings changes to the
evolution of scalar perturbation variables. This contrasts with the
fact that they do not interfere with each other in principle in standard
cosmology.

This paper consists as follows: In Section 2, we construct a class
of $f(R)$ theories satisfying the geometrical gauge symmetry of Weyl
geometry. By adopting an auxiliary scalar field and regarding the vector field as an electromagnetic potential, we establish a connection between the scalar field and the Weyl field. One can see thereby that the potential of the additional scalar field induces the dynamics of the Weyl field. Section 3 shows that the constructed model is equivalent
to Einstein gravity with two different scalar fields: One is
a scalar field coming from the modified gravitational action, and  the other may be deemed the Weyl field. They both exhibit non-minimal coupling valid only at the perturbational
scale. In Section 4, we study how the evolution of cosmological observables  in our modification differs from the standard model without primordial Weyl symmetry breaking, in large scales and
the first-order perturbation theory. As one interesting feature of the model, we found that there may be shifts in values of the Planck constant and the cosmological constant. In addition, when considering $f(R)$ inflation models, the existence of primordial GW also affects the evolution of scalar perturbation variables, not only of tensor perturbation variables.  Section 5 investigates
how the results from section 4 are reflected in the CMB anisotropy
observables with a case study on the Starobinsky inflation. Mainly, we examine how the ISW effect differs from Einstein's gravity by analyzing their cosmological impacts in detail.
Finally, we prove our claim that the primordial
Weyl symmetry breaking may give an incidental restriction in the decision
of the primordial GW amplitude when compared with observational
data. Section 6 summarizes our study and discusses its advantages
and drawbacks. We also suggest a few possible extensions of our research.

\section{$f(R)$ gravity from Weyl geometry}

To begin with, we first show how to construct $f(R)$ gravity theory satisfying Weyl
gauge symmetry. In the standard formalism of GR, a covariant derivative
$\mathbf{\nabla}$ is taken to be dependent on the metric tensor $g_{\mu\nu}$.
This dependency is called metric compatibility and is given by
the following relationship: 
\begin{equation}
\nabla_{\alpha}g_{\mu\nu}=0.
\end{equation}
From (1), one can obtain a usual expression of the connection $\mathbf{\Gamma}$
only using the metric tensor as follows: 
\begin{equation}
\Gamma_{\mu\nu}^{\alpha}=\frac{1}{2}g^{\alpha\gamma}(g_{\gamma\mu,\nu}+g_{\gamma\nu,\gamma}-g_{\mu\nu,\gamma}),
\end{equation}
which is known as the Levi-Civita connection. However, one may allow
more degrees of freedom to obtain a broader class of theories. One
of them we adopt here is the Weyl integrable geometry, which introduces
a new scalar field $\phi$ to define a new covariant derivative
$\bar{\nabla}$(or a new connection $\bar{\mathbf{\Gamma}}$): 
\begin{equation}
\bar{\nabla}_{\alpha}g_{\mu\nu}=g_{\mu\nu}\frac{\phi_{,\alpha}}{M_{\mathrm{P}}},
\end{equation}
where $M_{\mathrm{P}}$ is a Planck mass. One intriguing property
of (3) is that it manifests a new kind of symmetry. That is, (3) is
invariant under the following transformation: 
\begin{equation}
(g_{\mu\nu},\phi)\rightarrow(\Omega^{2}g_{\mu\nu},\phi+2M_{\mathrm{P}}\ln\Omega),
\end{equation}
where $\Omega$ is a gauge fixing parameter. Therefore, we can understand
(3) as a scaling symmetry associated with the scalar field $\phi$.
This symmetry is often called the Weyl gauge symmetry. We call the
scalar field $\phi$ giving Weyl gauge symmetry to spacetime a
Weyl scalar field. Now one finds that a volume form invariant under the transformation is $\mathrm{d}x^4 \sqrt{-g} e^{-2 \phi}$, where $g$ denotes a determinant of the metric. Similarly, a corresponding invariant scalar curvature is $e^{\phi} \mathcal{R}$, where $\mathcal{R}$ is a Ricci scalar computed with the new connection  $\bar{\mathbf{\Gamma}}$. Hence the standard Einstein-Hilbert action should be modified as follows:
\begin{equation}
S_\mathrm{EH}=\frac{M_{\mathrm{P}}^{2}}{2}\int\mathrm{d}x^{4}\sqrt{-g}e^{-2 \phi}\,e^{\phi}\mathcal{R}=\frac{M_{\mathrm{P}}^{2}}{2}\int\mathrm{d}x^{4}\sqrt{-g}\,e^{-\phi}\mathcal{R}.
\end{equation}
By the same token, we can easily write an  action with respect to the metric for an $f(R)$ gravity satisfying the Weyl gauge symmetry: 
\begin{equation}
S_g=\frac{M_{\mathrm{P}}^{2}}{2}\int\mathrm{d}x^{4}\sqrt{-g}e^{-2 \phi}\,f(e^{\phi}\mathcal{R}).
\end{equation}

However, one may claim that (6) is not a complete form of the action, as we can introduce the dynamics for the Weyl scalar field. For this reason, we now want to include an action for the Weyl scalar
field $\phi$. But herein follows a caution since it is tempting to write the entire action as follows: 
\begin{equation}
S=\frac{M_{\mathrm{P}}^{2}}{2}\int\mathrm{d}x^{4}\sqrt{-g}e^{-2 \phi}\,[f(e^{\phi}\mathcal{R})-\frac{1}{2}e^{\phi}g^{\mu\nu}\bar{\nabla}_{\mu}\phi\bar{\nabla}_{\nu}\phi-V(\phi)],
\end{equation}
where $V(\phi)$ is a potential for $\phi$. However, as \citep{1703_05649} correctly points out, this attempt is not correct because the kinetic term $e^{\phi}g^{\mu\nu}\bar{\nabla}_{\mu}\phi\bar{\nabla}_{\nu}\phi$ is not invariant under the transformation (4). Here we follow the discussion in the article above that we have just mentioned to tame this issue. By making use of the result from \citep{1703_05649}, we are forced to adopt one more additional vector field $B_\mu$ corresponding to a gauge boson for the Weyl geometry. Therefore, a fully gauge-invariant form of the action becomes
\begin{equation}
S=\frac{M_{\mathrm{P}}^{2}}{2}\int\mathrm{d}x^{4}\sqrt{-g}e^{-2 \phi}\,[f(e^{\phi}\mathcal{R})-\frac{1}{2}e^{\phi}g^{\mu\nu}\bar{\mathcal{D}}_{\mu}\phi\bar{\mathcal{D}}_{\nu}\phi-V(\phi)- \frac{1}{4}H_{\mu\nu}H^{\mu\nu} ],
\end{equation}
where $\bar{\mathcal{D}}_\mu\equiv\bar{\nabla}_\mu+\alpha B_\mu$  is a gauge covariant derivative, $H_{\mu\nu}\equiv \bar{\nabla}_\nu W_\mu - \bar{\nabla}_\mu W_\nu $ is a field strength tensor for $W_\mu\equiv\phi B_\mu/M_\mathrm{P}$, and $\alpha$ is a coupling constant. To ensure invariance in all respects, it is required that the following further transformation rules for the gauge boson and the potential hold,
\begin{equation}
(W_\mu,V(\phi))\rightarrow(W_\mu-2\alpha^{-1}\bar{\nabla}_\mu\ln\Omega,\bar{V}(\phi+2M_{\mathrm{P}}\ln\Omega)),
\end{equation}
where $\bar{V}$ is a specific form of the potential setting the action to be invariant. We call the action (8) a class of theories of Weyl $f(R)$ gravity.

Since the theory is invariant, one may choose a specific frame for one's
purpose. The frame we have used until now $(g_{\mu\nu},\phi)$
is called a Weyl frame. However, it is helpful to choose another frame
when we derive equations of motion. This new frame $(\gamma_{\mu\nu},0)$ with a gauge choose $\phi=-2M_{\mathrm{P}}\ln\Omega_0$,
where $\gamma_{\mu\nu}\equiv e^{-\phi/M_{\mathrm{P}}}g_{\mu\nu}$,
is called a Riemannian frame. One crucial property of the Riemannian
frame is that the connection $\bar{\mathbf{\Gamma}}$ behaves like
$\mathbf{\Gamma}$, i.e., the standard Levi-Civita connection. To
be explicit, we have an effective kind of metric compatibility that holds with the following modified metric: 
\begin{equation}
\nabla_{\alpha}\gamma_{\mu\nu}=0.
\end{equation}
This property is convenient when we conduct a computation
under a theory satisfying (4) since we can treat the derivative with
the rescaled metric $\gamma_{\mu\nu}$ just as we do in GR. So it is straightforward to see that (6) is exactly equivalent with
\begin{equation}
S_{\gamma}=\frac{M_{\mathrm{P}}^{2}}{2}\int\mathrm{d}x^{4}\sqrt{-\gamma}\,f(R_{\gamma}),
\end{equation}
where $R_{\gamma}$ is a Ricci scalar defined with the standard Levi-Civita connection. Though, one must note that all the connections and $R_{\gamma}$ should not be computed with respect to $g_{\mu\nu}$ but  $\gamma_{\mu\nu}$. By following the transformation rule, we obtain the action in the Riemannian frame as follows:
\begin{equation}
S=\frac{M_{\mathrm{P}}^{2}}{2}\int\mathrm{d}x^{4}\sqrt{-\gamma}\,[f(R_{\gamma})-\frac{1}{4}F_{\mu\nu}F^{\mu\nu} ],
\end{equation}
where we have assumed that $\bar{V}(0)=0$,  $F_{\mu\nu}\equiv \nabla_\nu A_\mu - \nabla_\mu A_\nu $ is a field strength tensor for $A_\mu$, and $A_\mu\equiv W_\mu-2\alpha^{-1}\nabla_\mu\ln\Omega_0=W_\mu+\alpha^{-1}M^{-1}_{\mathrm{P}}\nabla_\mu \phi$ is a redefined vector field. Here we want to emphasize that we let the electromagnetic field action survive, differing from the original discussion of \citep{1703_05649}. In the Riemannian frame, of course, the motivation for why we have adopted the gauge field becomes obsolete. However, it does not obstruct us from abandoning it. In fact, it seems that there is no clue for us to set $A_{\mu}=0$, at least from the transformation rules (9). Nevertheless, if we introduce an auxiliary scalar field and its potential $\bar{\phi}$ and $V_{\bar{\phi}}(\bar{\phi})$, one may suggest that $A_\mu$ acts like the ordinary electromagnetic potential. If this is the case, the action becomes
\begin{equation}
S=\frac{M_{\mathrm{P}}^{2}}{2}\int\mathrm{d}x^{4}\sqrt{-\gamma}\,[f(R_{\gamma})-\frac{1}{2}\gamma^{\mu\nu}\mathcal{D}_{\mu}\bar{\phi}\mathcal{D}_{\nu}\bar{\phi}-V_{\bar{\phi}}(\bar{\phi})- \frac{1}{4}F_{\mu\nu}F^{\mu\nu} ],
\end{equation}
where we have defined  $\mathcal{D}_\mu\equiv\nabla_\mu+\alpha A_\mu$  a Riemannian gauge covariant derivative. If we put $\bar{\phi}=M_\mathrm{P}$ and $V_{\bar{\phi}}(M_\mathrm{P})=\bar{V}(0)=0$, this is equivalent to (12). However, writing out an action for $\bar{\phi}$ as if it has dynamical values turns out to be helpful. To clarify what we mean, we want to pass over the function of $A_\mu$ to the new scalar field. If we  require  $A_\mu$  to play a role as an electromagnetic potential, one must impose the following transformation rules similar to (4) so that (13) is left invariant thereunder:
\begin{equation}
(\bar{\phi},A_\mu,V_{\bar{\phi}}(\bar{\phi}))\rightarrow(\bar{\phi}\varphi/M_\mathrm{P},A_\mu-\alpha^{-1}\nabla_\mu\ln{(\varphi/M_\mathrm{P})},\hat{V}_{\bar{\phi}}(\bar{\phi}\varphi/M_\mathrm{P})),
\end{equation}
where $\varphi$ is a gauge fixing parameter, and $\hat{V}_{\bar{\phi}}$ is a modified potential to obey the invariance. If we now pick a specific gauge so that $\nabla_\mu \ln{(\varphi/M_\mathrm{P})}=\alpha A_\mu$, we can safely neglect terms on $A_\mu$. Now putting $\bar{\phi}=M_\mathrm{P}$ back again, we can encapsulate our discussion as the following action:
\begin{equation}
S=\frac{M_{\mathrm{P}}^{2}}{2}\int\mathrm{d}x^{4}\sqrt{-\gamma}\,[f(R_{\gamma})-\frac{1}{2}\gamma^{\mu\nu}\mathcal{D}_{\mu}\varphi\mathcal{D}_{\nu}\varphi-V_{\varphi}(\varphi)],
\end{equation}
where we have set $V_{\varphi}(\varphi)\equiv\hat{V}_{\bar{\phi}}(\varphi)$. Hence, we may deem the gauge fixing parameter $\varphi$ to behave like the dynamical field in this frame. Furthermore, since $\varphi$ is related to the Weyl field $\phi$ with gauge conditions, we can expect that the dynamics of $\varphi$ would also bring on the evolution of the Weyl field. By taking the computational steps from now on reversely, one can recover the original theory (8) from (15). From this context, we may say that the Weyl symmetry is hidden but could be manipulated with the help of  $V_{\varphi}(\varphi)$. This feature will be elaborated on in Section 3 in detail.

For completeness, we close this section by finding equations of motion of (15). Taking a variation with the rescaled metric and the Weyl scalar
field, we obtain the following equations for the metric and the Weyl scalar
field: 
\begin{align}
 & (R_\gamma)_{\mu\nu} F(R_\gamma) -\frac{1}{2}R_\gamma \gamma_{\mu\nu}-\mathcal{D}_{\mu}\mathcal{D}_{\nu}F(R_\gamma)+\gamma_{\mu\nu}\square F(R_\gamma)\nonumber \\
 & =\kappa T_{\mu\nu}+T_{\mu\nu}^{(\varphi)},
\end{align}
\begin{equation}
\square \varphi=\frac{\mathrm{d}V_{\varphi}}{\mathrm{d}\varphi},
\end{equation}
where $F\equiv\partial f/\partial\mathcal{R}$, $\square\equiv\gamma^{\mu\nu}\mathcal{D}_\mu \mathcal{D}_\nu$ is the D’Alambertian operator, $T_{\mu\nu}$ is an
energy-momentum(EM) tensor for ordinary matter, including the cosmological
constant, and the EM tensor for the Weyl field in the Riemannian frame $T_{\mu\nu}^{(\varphi)}$ is
defined by 
\begin{equation}
T_{\mu\nu}^{(\varphi)}\equiv\frac{1}{2}[\mathcal{D}_{\mu}\varphi\mathcal{D}_{\nu}\varphi-\frac{1}{2}\gamma_{\mu\nu}\{\gamma^{\alpha\beta}\mathcal{D}_{\alpha}\varphi\mathcal{D}_{\beta}\varphi+V_{\varphi}(\varphi)\}].
\end{equation}

\section{Equivalence with two non-minimally coupled scalar fields}

From now on, we will work only in the Riemannian frame with a gauge choice so that the vector field vanishes. For this reason, we write $\nabla_\mu$ instead of $\mathcal{D}_{\mu}$ for the covariant derivative with no worry about confusion.  To invoke primordial symmetry
breaking of $\varphi$ at the Planck scale, we adopt the following Higgs-type potential: 
\begin{equation}
V_{\varphi}(\varphi)=V_{0}(\varphi^{2}-M_{\mathrm{P}}^{2})^{2}.
\end{equation}
To explain how the symmetry breaking of $\varphi$ is connected to the Weyl gauge symmetry breaking, let us recall the relationship  between the electromagnetic potential and scalar fields: $\alpha A_\mu=\nabla_\mu \ln{(\varphi/M_\mathrm{P})}=\alpha W_\mu+\nabla_\mu (\phi/M_\mathrm{P})$. If we assume the coupling constant $\alpha$ is small enough, we have $\varphi\approx M_\mathrm{P}e^{\phi/M_\mathrm{P}}$. Thus, we claim that the potential (19) also naturally induces the Weyl symmetry breaking, not only the symmetry of $\varphi$, by bringing down the symmetry (4) from all scales to perturbational scales only. And by the same logic, we will from now on work with $\varphi$ rather than $\phi$  and call it the Weyl field. After the symmetry breaking, we expand $\varphi=M_{\mathrm{P}}(1+\delta\varphi)$
where $\delta\varphi$ is a perturbation of $\varphi$. In the same
manner, $R_\gamma$ can be expanded up to the first order of $\delta\varphi$
as follows: 
\begin{equation}
R_\gamma=\gamma^{\mu\nu}(R_g)_{\mu\nu}-3 \square\delta\varphi,
\end{equation}
where $(R_g)_{\mu\nu}$ denotes the Ricci tensor with the Levi-Civita connection with respect to the original metric $g_{\mu\nu}$. Note that we have exploited the fact that  $\gamma_{\mu\nu}=e^{-\phi/M_{\mathrm{P}}}g_{\mu\nu}\approx (1-\delta\varphi)g_{\mu\nu}$ when computing the connection and the Ricci tensor. However, the reader should be aware that we regard $\gamma_{\mu\nu}$ as fixed and not perturbed
by the expansion of $\varphi$ so as to exploit the metric compatibility
(10) in the Riemannian frame. Moreover, after the symmetry breaking, the equation of motion for $\delta\varphi$ becomes 
\begin{equation}
\square\delta\varphi\approx m_{\varphi}^{2}\delta\varphi,
\end{equation}
where $m_{\varphi}^{2}\equiv4V_{0}M_{\mathrm{P}}^{2}$. Here one should
note that this is not an exact form of the equation of motion for
$\delta\varphi$, as we did not consider the effect of the symmetry
breaking in the gravity action yet. However, we will use this equation
as an approximation when we transform the action as a standard gravity
with a non-minimally coupled field at the non-perturbative scale.
This assumption is, of course, not the fully exact way to compute
the action. However, we will see that this has some benefits as we
may reduce the high-order derivative part of the action.

Our goal in this section is to find an equivalent form of (19) whose gravitational sector of the
action is the Einstein-Hilbert action with a non-minimally coupled scalar field. The effect of the modified action will be reflected in the potential of the scalar field.
To find such a theory, let us first review how one can find an equivalent
theory with the scalar field in general $f(R)$ theories without symmetry
breaking, following the methodology of \citep{23}. We claim that
the following action is equivalent to the action of $f(R)$ theory,
that is, $S_{g}=\int\mathrm{d}^{4}x\sqrt{-g}\,(M_{\mathrm{P}}^{2}/2)f(R)$:
\begin{equation}
S_{g}=\int\mathrm{d}^{4}x\sqrt{-g}\,\frac{M_{\mathrm{P}}^{2}}{2}[f(\chi)+f^{\prime}(\chi)(R-\chi)],
\end{equation}
where we have adopted an auxiliary field $\chi$. Varying (22) with
respect to the new field $\chi$, one finds that $f^{\prime\prime}(\chi)(R-\chi)=0$
so that $\chi=R$ and our claim is proved provided that $f^{\prime\prime}(\chi)\neq0$.
If we define a new field $\zeta\equiv f^{\prime}(\chi)$, one may
rewrite (22) as 
\begin{equation}
S_{g}=\int\mathrm{d}x^{4}\sqrt{-g}\,[\frac{M_{\mathrm{P}}^{2}}{2}\zeta R-U(\zeta)],
\end{equation}
where the potential $U(\zeta)$ is given by 
\begin{equation}
U(\zeta)\equiv\frac{M_{\mathrm{P}}^{2}}{2}[\chi(\zeta)\zeta-f(\chi(\zeta))].
\end{equation}
Now the action (23) can be expressed as a non-minimally coupled one
when adopting an appropriate metric rescaling. However, when one tries
to find an equivalent action in the presence of Weyl symmetry breaking,
one needs to pay more attention because the newly defined field $\zeta$
is clearly associated with the perturbative terms of $\delta\varphi$.
Here we want to restrict that the non-minimally coupled scalar field
only comes from the non-perturbative part of $f(R_\gamma)$ to see new effects from the symmetry breaking more clearly.
So it would be of some comfort to abstract the perturbative part from
the original action. To this end, we extract the gravity part of the
action (19) up to the first order of $\delta\varphi$. we obtain 
\begin{eqnarray}
S_{\gamma} & = & \int\mathrm{d}^{4}x\sqrt{-\gamma}\,\frac{M_{\mathrm{P}}^{2}}{2}f(\gamma^{\mu\nu}(R_g)_{\mu\nu}-3\square\delta\varphi)\nonumber \\
 & = & \int\mathrm{d}^{4}x\sqrt{-\gamma}\,\frac{M_{\mathrm{P}}^{2}}{2}f(\gamma^{\mu\nu}(R_g)_{\mu\nu})\nonumber \\
 &  & -\int\mathrm{d}^{4}x\sqrt{-\gamma}\,\frac{M_{\mathrm{P}}^{2}}{2}[3f^{\prime}(\gamma^{\mu\nu}(R_g)_{\mu\nu})m_{\delta\varphi}^{2}\delta\varphi].
\end{eqnarray}
Here we used the relation (21) to reduce the high derivative order
of the field $\delta\varphi$. The readers should be aware that we
have expanded the action only up to the first order of the perturbed
Weyl field. To be consistent with the procedure
when we compute a perturbation of the action, we would have to consider
up to the second order of $\delta\varphi$. However, since our current
goal is merely to consider the effect of the primordial symmetry breaking
in the $f(R)$ theory, we expect that only considering up
to the first order in the function $f$ would be enough for our current
purpose. Note that we suppress the Weyl field part of the action and
only consider $S_{\gamma}$ for the moment for consistency and convenience.

From this expression, we introduce a new scalar field that only comes
from the non-perturbative part of $f(R_\gamma)$: 
\begin{equation}
\chi\equiv f^{\prime}(\gamma^{\mu\nu}(R_g)_{\mu\nu}).
\end{equation}
Proceeding similar steps as we saw from (22) to (24), we express (25) as
\begin{equation}
S_{\gamma}=\int\mathrm{d}^{4}x\sqrt{-\gamma}\,[\frac{M_{\mathrm{P}}^{2}}{2}(\chi+\psi)R_\gamma-U(\chi,\psi)],
\end{equation}
where we have defined 
\begin{equation}
\psi\equiv-3m_{\delta\varphi}^{2}f^{\prime\prime}(\gamma^{\mu\nu}(R_g)_{\mu\nu})\delta\varphi,
\end{equation}
and the potential is given by 
\begin{equation}
U(\chi,\psi)=\frac{M_{\mathrm{P}}^{2}}{2}[(\chi+\psi)R_\gamma-f(R_\gamma)].
\end{equation}
To clarify the effect of the Weyl symmetry breaking on the potential,
we split the potential as $U(\chi,\psi)=U_{\chi}(\chi)+\delta U(\chi,\psi)$,
where $\delta U$ is valid only at the perturbative scale. Expanding
(29) with respect to $\delta\varphi$ to the first order at most, we obtain the
following: 
\begin{align}
U_{\chi}(\chi) & =\frac{M_{\mathrm{P}}^{2}}{2}[\chi\gamma^{\mu\nu}(R_g)_{\mu\nu}-f(\gamma^{\mu\nu}(R_g)_{\mu\nu})]\nonumber \\
 & =\frac{M_{\mathrm{P}}^{2}}{2}[\chi(f^{\prime})^{-1}(\chi)-f((f^{\prime})^{-1}(\chi))]
\end{align}
\begin{align}
\delta U(\chi,\psi) & =\frac{M_{\mathrm{P}}^{2}}{2}\psi\gamma^{\mu\nu}(R_g)_{\mu\nu}\nonumber \\
 & =\frac{M_{\mathrm{P}}^{2}}{2}\psi(f^{\prime})^{-1}(\chi).
\end{align}
Now we may perform a conformal transformation to find an action that
is non-minimally coupled with the scalar field $\chi$ at the non-perturbative
scale. First, let us define the rescaled metric 
\begin{equation}
\hat{\gamma}_{\mu\nu}=\chi\gamma_{\mu\nu}.
\end{equation}
Note that this transformation does not correspond to the gauge transformation
(4) and is merely a technique for obtaining the non-minimally coupled action.
Rewriting our action (27) with respect to the newly defined metric
(32), we finally arrive at the following result: 
\begin{align}
S_{\hat{\gamma}}=\int\mathrm{d}^{4}x\sqrt{-\hat{\gamma}}\,[ & \frac{M_{\mathrm{P}}^{2}}{2}(1+\psi e^{-\sqrt{2/3}\zeta/M_{\mathrm{P}}})R_{\hat{\gamma}}\nonumber \\
 & -\frac{1}{2}\hat{\gamma}^{\mu\nu}\hat{\nabla}_{\mu}\zeta\hat{\nabla}_{\nu}\zeta-V(\zeta)-\delta V(\zeta,\psi)],
\end{align}
where the terms with hat denote that it is with respect to the rescaled
metric $\hat{\gamma}_{\mu\nu}$, $\zeta/M_{\mathrm{P}}\equiv\sqrt{3/2}\ln\chi$,
and 
\begin{equation}
V(\zeta)\equiv e^{-2\sqrt{2/3}\zeta/M_{\mathrm{P}}}U(e^{\sqrt{2/3}\zeta/M_{\mathrm{P}}}),
\end{equation}
\begin{equation}
\delta V(\zeta,\psi)\equiv e^{-2\sqrt{2/3}\zeta/M_{\mathrm{P}}}\delta U(e^{\sqrt{2/3}\zeta/M_{\mathrm{P}}},\psi).
\end{equation}
Hence, the effect of Weyl symmetry breaking is manifested as an additional
non-minimal coupling only valid at the perturbative scale. Recovering
the suppressed action for $\varphi$, we write the total action with
an action for ordinary matter $S_{\mathrm{M}}$ as follows:
\begin{align}
S_{\hat{\gamma}}=\int\mathrm{d}^{4}x\sqrt{-\hat{\gamma}}\,[ & \frac{M_{\mathrm{P}}^{2}}{2}(1-3m_{\delta\varphi}^{2}f^{\prime\prime}((f^{\prime})^{-1}(e^{\sqrt{2/3}\zeta/M_{\mathrm{P}}}))e^{-\sqrt{2/3}\zeta/M_{\mathrm{P}}}\delta\varphi)\hat{\mathcal{R}}\nonumber \\
 & -\frac{1}{2}\hat{\gamma}^{\mu\nu}\hat{\nabla}_{\mu}\zeta\hat{\nabla}_{\nu}\zeta-V(\zeta)\nonumber \\
 & -\frac{1}{2}\gamma^{\mu\nu}\hat{\nabla}_{\mu}\delta\varphi\hat{\nabla}_{\nu}\delta\varphi-\frac{1}{2}m_{\varphi}^{2}\delta\varphi^{2}\nonumber \\
 & -\delta V(\zeta,-3m_{\delta\varphi}^{2}f^{\prime\prime}((f^{\prime})^{-1}(e^{\sqrt{2/3}\zeta/M_{\mathrm{P}}}))\delta\varphi)]+S_{\mathrm{M}}.
\end{align}

\section{Two effects of symmetry breaking}

To clarify how our modification differs from the original $f(R)$
gravity, we study how cosmological evolution differs from one
in standard cosmology. To this end, we want to force the field
$\phi$ to be suppressed. We take this crude approach
to emphasize phenomena that may differ from $\Lambda$CDM cosmology
and to distinguish them from the effect of specific $f(R)$ models.
This enables us to guarantee that the other phenomena we seek from
now on purely originate from the result of Weyl symmetry breaking. That is,
one may neglect effects from modifying the gravitational
action, i.e., the Einstein-Hilbert action to the $f(R)$ form at large
or nonperturbative scales. Nevertheless, we leave the Weyl field part
as it is so we can observe the effects of the symmetry breaking. Also,
the potential for the Weyl field depends on the form of the function
$f(R)$. So  one still can distinguish what $f(R)$ model is chosen in
the first place.

From the discussion above, we assume that the field $\zeta$ is stabilized
enough to hide the scalar field $\zeta$ part of the action. To be more concrete, we suppress the kinetic term of the scalar
field, that is, $\left|\hat{\gamma}^{\mu\nu}\hat{\nabla}_{\mu}\zeta\hat{\nabla}_{\nu}\zeta\right|\ll\left|V(\zeta)\right|$,
even though it must be regarded as a dynamical variable as it evolves
through time. Furthermore, we assume that the field is stabilized
to a local(or global) minimum of the potential. Say $V_{*}=V(\zeta_{*})$
as such a value of the potential, where $\zeta_{*}$ is a stable attracting
point of the field. Now the action (29) can be read as
\begin{align}
S_{\hat{\gamma}}=\int\mathrm{d}^{4}x\sqrt{-\hat{\gamma}}\,[ & \frac{M_{\mathrm{P}}^{2}}{2}\{1-m_{\delta\varphi}^{2}(C_{1}\delta\varphi_{*}+C_{2})\} R_{\hat{\gamma}} -V_{*}\nonumber \\
 & -\frac{1}{2}\gamma^{\mu\nu}\hat{\nabla}_{\mu}\delta\varphi_{*}\hat{\nabla}_{\nu}\delta\varphi_{*}-\frac{1}{2}m_{\varphi}^{2}\delta\varphi_{*}^{2}]+S_{\mathrm{M}},
\end{align}
where $\delta\varphi_{*}$ is defined so that $V(\delta\varphi_{*})=m_{\varphi}^{2}\delta\varphi_{*}^{2}/2$,
and the coefficients $C_{1}$ and $C_{2}$ depends on the $f(R)$
model. To be more specific, the change in the minimum the potential
$V(\delta\varphi)$, which derives the coefficient $C_{2}$, comes
from the $\delta V$. If $\zeta_{*}$ has a finite value, there would
be a possibility that $C_{2}\neq0$. Otherwise, if $\zeta_{*}=\infty$,
it is likely to be zero as it would be suppressed by the exponential
factor $e^{-2\sqrt{2/3}\zeta/M_{\mathrm{P}}}$ in (35). For many $f(R)$
models with no transcendental function of $R$, the corresponding
field is likely to be stabilized as $\zeta\rightarrow\infty$. In this
case, the dominating term in the potential is the exponential one,
that is, $V(\zeta)\approx V_{0}e^{-\alpha_{0}\sqrt{2/3}\zeta/M_{\mathrm{P}}}$,
where $V_{0}$ and $\alpha_{0}$ are some constants. If one supposes that
the selected $f(R)$ model should play a role as the dark energy,
it can be deemed that, in this case, the scalar field $\zeta$ acts as
a quintessence. In a way, one may suspect that many $f(R)$ dark
energy models would behave similarly to quintessence models with exponential
potential. Furthermore, for this case, $V_{*}$ may be regarded as $\Lambda$
in the standard scenario, so the field perfectly mimics the cosmological
constant. 

Let us clarify the meaning of these coefficients by studying equations
of motion. Up to the first-order perturbation, we find the equations of
motion:
\begin{equation}
\{1-m_{\delta\varphi}^{2}(C_{1}\delta\varphi_{*}+C_{2})\}(G_{\hat{\gamma}})_{\mu\nu}=\kappa T_{\mu\nu}^{M}+\kappa T_{\mu\nu}^{(\delta\varphi_{*})},
\end{equation}
\begin{equation}
\hat{\square}\delta\varphi_{*}=m_{\varphi}^{2}(\delta\varphi_{*}+\frac{C_{1}M_{\mathrm{P}}^{2}}{2}\delta R_{\hat{\gamma}}),
\end{equation}
where $\kappa\equiv2/M_{\mathrm{P}}^{2}$, $T_{\mu\nu}^{M}$ is the
energy-momentum tensor for ordinary matter with $V_{*}$ or the dark
energy, $\delta R_{\hat{\gamma}}$ is a perturbed Ricci tensor and
\begin{align}
T_{\mu\nu}^{(\delta\varphi_{*})}\equiv\hat{\nabla}_{\mu}\delta\varphi_{*}\hat{\nabla}_{\nu}\delta\varphi_{*}-\hat{\gamma}_{\mu\nu}[\frac{1}{2}\hat{\gamma}^{\alpha\beta}\hat{\nabla}_{\mu}\delta\varphi_{*}\hat{\nabla}_{\nu}\delta\varphi_{*}+\frac{1}{2}m_{\varphi}^{2}\delta\varphi_{*}^{2}]\nonumber \\
-C_{1}m_{\delta\varphi}^{2}(\hat{\nabla}_{\mu}\hat{\nabla}_{\nu}-\hat{\gamma}_{\mu\nu}\hat{\square})\delta\varphi_{*}.
\end{align}
Now one easily finds a genuine change at the nonperturbative
level when considering $C_{2}$. Rewriting (38) by ignoring perturbative
terms, the Einstein equation is changed as follows:
\begin{equation}
(1-m_{\delta\varphi}^{2}C_{2})(G_{\hat{\gamma}})_{\mu\nu}=\kappa T_{\mu\nu}^{M}.
\end{equation}
One may think that this corresponds to the effective change of the
the cosmological constant and the Planck mass are given by
\begin{equation}
\Lambda\rightarrow\Lambda^{*}\equiv\frac{\Lambda}{(1-m_{\delta\varphi}^{2}C_{2})},
\end{equation}
\begin{equation}
M_{\mathrm{P}}\rightarrow M_{\mathrm{P}}^{*}\equiv(1-m_{\delta\varphi}^{2}C_{2})^{1/2}M_{\mathrm{P}}.
\end{equation}

Hence we arrive at our first important observation. Even though they
are not perturbational variables, as this shift of values of the cosmological
constant and the Planck mass is only viable by considering perturbational
theory, this effect could be viewed as a new example of backreaction
effects. Of course, there is no clue that the Weyl symmetry breaking
really happened. Nevertheless, we believe that this result
might be one of many reasons to consider backreaction in the early universe.

Primarily, many researches on the $f(R)$ gravity theory want to supply novel alternatives
to the $\Lambda$CDM model that may give us a better explanation of cosmological problems.  In contrast with these studies, we would especially like to emphasize that our model
suggests neither a new model for dark energy nor an unnecessary modification
of the background equations. Of course, there must be additional changes
when adopting dark energy models differing from a cosmological constant.
However, when the equation of state for the dark energy model converges
to $w\approx-1$ enough, we expect that one could obtain almost the
same result independent of the models except for the value of $C_{2}$. On top of that, it is presumably natural to consider high-curvature correction
to the Einstein-Hilbert action from the motivation of quantum gravity.
Also, symmetry-breaking phenomena are universal in physics, and
there is no reason not to consider them also in gravity. Hence, it
would be quite conceivable to imagine this kind of scenario or at
least a similar one in studying quantum gravity. Additionally, many
experiments are already out there in which constraints
allowed values of the gravitational constant. For this reason, we
suggest a need for a much deeper investigation without
the crude approximation. However, this will require a further detailed
study beyond this manuscript's current scope, and we leave this
as a future study topic. By now, we suspect that there would be not
so much room for $f(R)$ models with nonzero values of $C_{2}$ allowing
primordial Weyl symmetry breaking, as more restrictive experimental
data are coming out nowadays.

Now let us focus on the perturbational regime and its consequences
on CMB anisotropy. From now on, we assume that a specific $f(R)$ model is suggested to give inflation in the universe. Also, from now on, we set  $C_{2}$ as zero
or has a very close value to zero, at least. This condition can be
justified if there appears no higher order terms than $R^{2}$ when
expanding $f(R)$ with polynomials because $C_{2}$ basically comes
from the value of $f^{\prime\prime}$. Now we consider what
function $C_{1}$ has. Let us now write the perturbed Einstein equation
explicitly:
\begin{equation}
\delta (G_{\hat{\gamma}})_{\mu\nu}=\kappa\delta T_{\mu\nu}^{M}-C_{1}m_{\delta\varphi}^{2}\{(\hat{\nabla}_{\mu}\hat{\nabla}_{\nu}-\hat{\gamma}_{\mu\nu}\hat{\square})\delta\varphi-\left\langle T_{\mu\nu}^{M}\right\rangle \delta\varphi\},
\end{equation}
where $\left\langle \right\rangle $ denotes an averaged value. From
this equation, we find that the model approaches GR as $m_{\delta\varphi}\rightarrow0$
and $C_{1}$ plays a role in amplifying an oscillation of $\delta\varphi$,
so that gives more metric fluctuations; besides, $C_{1}$ also has
a meaning that could rival the meaning of $C_{2}$,
by showing an imprint of the tensor perturbation on the evolution
of scalar perturbation variables when one considers an $f(R)$ theory
as an inflationary model. The reason why is that the initial amplitude
for primordial GW $\mathcal{P}_{\mathrm{T}}$ arising from the inflation
depends on a form of the inflationary potential $V(\phi)$, that is,
the form of the function $f(R)$. 

Thus, we arrive at our second important
conclusion: the coefficient $C_{1}$ is a function of $\mathcal{P}_{\mathrm{T}}$.
And since this genuine feature is independent of the type of $f(R)$
theories, the primordial Weyl symmetry breaking must leave some footprints
of primordial GW during the evolution of scalar variables. In fact,
this modification only affects the scalar part, as it comes from a
new perturbative scalar field. There is no additional source for GW; their behavior merely depends on what type of inflation the
given $f(R)$ model provides. From the discussion, we conclude that we
need to dive into how CMB anisotropies change in $f(R)$ inflationary
models with the Weyl symmetry breaking. This new phenomenon is impossible
in the standard cosmology in principle by the orthogonality between
the spherical harmonics for scalar and tensor, which we adopt when
we conduct the scalar-vector-tensor decomposition of perturbational
variables.

\section{Case study: Starobinsky inflation}

In this section, we want to investigate the imprint of the tensor
perturbation on the CMB scalar anisotropy with an example of $f(R)$
gravity models, which is known as Starobinsky inflation \citep{25}, whose function
$f$ is given by 
\begin{equation}
f(R)=R+\frac{R^{2}}{6M^{2}},
\end{equation}
where $M$ is a mass scale constant. From now on, $R$ means $R_{\gamma}$ always, and we denote $\hat{R}\equiv R_{\hat{\gamma}}$ and so on. Although many modifications of Starobinsky inflation are already out there, we believe that it would
suffice to study only the original Starobinsky model, as (45) has
terms only up to the second order of the Ricci tensor, i.e., $C_{2}=0$.
Moreover, we expect that most of the modifications would exhibit
no further extraordinary features except for what we have already shown
in the previous paragraph. For a detailed study, we now repeat the
steps we have previously discussed: first, the Weyl gauge symmetry
becomes broken at the Planck energy scale due to the Higgs-type potential.
After the symmetry breaking, the transformation formula (4) is only
valid at the perturbative scale, and the Weyl scalar field becomes
a perturbative variable. At non-perturbative scales, the potential
$V(\zeta)$ is known as Starobinsky potential: 
\begin{equation}
V(\zeta)=\frac{3}{4}M^{2}M_{\mathrm{P}}^{2}(1-e^{-\sqrt{2/3}\zeta/M_{\mathrm{P}}})^{2}.
\end{equation}
For the Starobinsky model, we find that 
\begin{equation}
\psi=-(\frac{m_{\varphi}}{M})^{2}\delta\varphi,
\end{equation}
\begin{equation}
\delta V(\zeta,\delta\varphi)\equiv\frac{3}{2}m_{\varphi}^{2}M_{\mathrm{P}}^{2}e^{-2\sqrt{2/3}\zeta/M_{\mathrm{P}}}(1-e^{\sqrt{2/3}\zeta/M_{\mathrm{P}}})\delta\varphi.
\end{equation}
From now on, let us restrict our interest to the first-order perturbation
theory. Varying the action (36) with $\hat{\gamma}_{\mu\nu}$ and
$\zeta$, we obtain equations of motion: 
\begin{equation}
\{1+\psi e^{-\sqrt{2/3}\zeta/M_{\mathrm{P}}}\}\hat{G}_{\mu\nu}=\kappa T_{\mu\nu}^{(\zeta)}+T_{\mu\nu}^{\mathrm{int}},
\end{equation}
\begin{equation}
\hat{\square}\zeta=\partial_{\zeta}[V(\zeta)+\delta V(\zeta,\psi)],
\end{equation}
where $\kappa\equiv2/M_{\mathrm{p}}^{2}$ and 
\begin{equation}
T_{\mu\nu}^{(\zeta)}\equiv\hat{\nabla}_{\mu}\zeta\hat{\nabla}_{\nu}\zeta-\hat{\gamma}_{\mu\nu}[\frac{1}{2}\hat{\gamma}^{\alpha\beta}\hat{\nabla}_{\mu}\zeta\hat{\nabla}_{\nu}\zeta+V(\zeta)],
\end{equation}
\begin{align}
T_{\mu\nu}^{\mathrm{int}}\equiv-(\frac{m_{\varphi}}{M})^{2}(\hat{\nabla}_{\mu}\hat{\nabla}_{\nu}-\hat{\gamma}_{\mu\nu}\hat{\square})(e^{-\sqrt{2/3}\zeta/M_{\mathrm{P}}}\delta\varphi)\nonumber \\
-\hat{\gamma}_{\mu\nu}\delta V(\zeta,\psi),
\end{align}
where $T_{\mu\nu}^{\mathrm{int}}$ denotes the interaction between
$\delta\varphi$ and $\zeta$. One might think that the additional
terms may affect the perturbation of the inflaton field $\zeta$. However,
we can fix $\delta\varphi=\delta\zeta=0$ if we adopt the following
gauge: 
\begin{equation}
\hat{\gamma}_{\mu\nu}=-\mathrm{d}t^{2}+a(t)^{2}\mathrm{d}x^{2}[(1-2\mathcal{A})\delta_{ij}+h_{ij}],
\end{equation}
where $\mathcal{A}$ and $h_{ij}$ are gauge variables satisfying $\partial_{i}h^{ij}=h_{i}^{\phantom{i}i}=0$. Hence one
can prove that there is no effect from Weyl gauge symmetry breaking
on quantum fluctuations during inflation.

After the inflation, the gravity sector of the action becomes 
\begin{equation}
S_{\gamma}=\int\mathrm{d}^{4}x\sqrt{-\gamma}\,[\frac{M_{\mathrm{P}}^{2}}{2}\{1-(\frac{m_{\varphi}}{M})^{2}\delta\varphi\}R].
\end{equation}
Here one must note that we may use the original metric before the
conformal transformation at the first order scale since $\chi=e^{\sqrt{2/3}\zeta/M_{\mathrm{P}}}\approx1$,
so $\hat{\gamma}_{\mu\nu}\approx\gamma_{\mu\nu}$. From the action
(54), we can see that the non-minimal coupling of the Weyl field $\delta\varphi$
and the scalar curvature still survives. Furthermore, as we have mentioned,
this coupling gives us one of the crucial imprints of the inflation
theory when one recalls that the mass $M$ inside the coupling constant
is related to the tensor power spectrum $\mathcal{P}_{\mathrm{T}}$
by the result of the Starobinsky inflation \citep{23}: 
\begin{equation}
\mathcal{P}_{\mathrm{T}}\approx\frac{4}{\pi}(\frac{M}{M_{\mathrm{P}}})^{2}.
\end{equation}
Hence, it is evident that the existence of primordial GW affects the results, especially for scalar perturbation, as
we expected. This feature clearly distinguishes the effect of Weyl
symmetry breaking in Starobinsky gravity from our previous result
\citep{21}, wherein the results cannot be associated with the inflation
theory. Furthermore, since a smaller amplitude of GW requires a stronger
non-minimal coupling, one may expect that our model poses a more considerable
minimum value for a scalar-to-tensor ratio than in standard $\Lambda$CDM
cosmology when compared with observational data. We will investigate
this issue deeper in later.

Now let us bring the Weyl field action again to observe how this affects the cosmological observables.
\begin{align}
S=\int\mathrm{d}^{4}x\sqrt{-\gamma}\,[\frac{M_{\mathrm{P}}^{2}}{2}\{1-(\frac{m_{\varphi}}{M})^{2}\delta\varphi\}R\nonumber \\
-\frac{1}{2}\gamma^{\mu\nu}\nabla_{\mu}\delta\varphi\nabla_{\nu}\delta\varphi-\frac{1}{2}m_{\varphi}^{2}\delta\varphi^{2}]\nonumber \\
+S_{\mathrm{M}},
\end{align}
where $S_{\mathrm{M}}$ is the action for ordinary matters, including
matters coming from the fluctuation during inflation. Up to the first
order of $\delta\varphi$, we obtain the equation of motion for the
metric. 
\begin{align}
[1-(\frac{m_{\varphi}}{M})^{2}\delta\varphi]G_{\mu\nu}=\kappa T_{\mu\nu}- & (\frac{m_{\varphi}}{M})^{2}(\nabla_{\mu}\nabla_{\nu}-\gamma_{\mu\nu}\square)\delta\varphi,
\end{align}
where $G_{\mu\nu}\equiv R_{\mu\nu}-R\gamma_{\mu\nu}/2$
is the Einstein tensor and $T_{\mu\nu}^{(\mathrm{M})}$
is the energy-momentum tensor of ordinary matters. Meanwhile, one
must be careful when one tries to derive an equation of motion for
$\delta\varphi$ since it is only valid at the perturbative level.
That is, the gravity action coupled with $\delta\varphi$ is not the
whole part of the scalar curvature but only its first-order perturbative
part, namely $\delta R$. Hence, the new equation of motion for $\delta\varphi$
is given as follows: 
\begin{equation}
\square\delta\varphi=m_{\varphi}^{2}[\delta\varphi+M^{-2}\delta R].
\end{equation}
Now, let us find the complete form of the Einstein equation at the
perturbative scale. We expand the Einstein tensor up to the first
order of $\delta\varphi$: 
\begin{equation}
G_{\mu\nu}=\mathcal{G}_{\mu\nu}-(\nabla_{\mu}\nabla_{\nu}\delta\varphi-\gamma_{\mu\nu}\square\delta\varphi),
\end{equation}
where $\mathcal{G}_{\mu\nu}$ is the Einstein tensor with respect to
the original metric $g_{\mu\nu}$. Furthermore, since we only consider
the first-order theory, all the derivatives, such as $\nabla_{\alpha}\delta\varphi$
and $\square\delta\varphi$, are now reduced to be also defined with  $g_{\mu\nu}$. Finally, we
obtain the following results: 
\begin{align}
\delta \mathcal{G}_{\mu\nu}=\kappa\delta T_{\mu\nu}^{(\mathrm{M})}+ & \kappa(\frac{m_{\varphi}}{M})^{2}\left\langle T_{\mu\nu}^{(\mathrm{M})}\right\rangle \delta\varphi\nonumber \\
 & +[1-(\frac{m_{\varphi}}{M})^{2}](\nabla_{\mu}\nabla_{\nu}-g_{\mu\nu}\square)\delta\varphi,
\end{align}
where $\left\langle T_{\mu\nu}^{(\mathrm{M})}\right\rangle $ is the average value of the energy-momentum tensor. Taking a Trace of (60)
to find $\delta R$ and substituting it into (58), we have 
\begin{align}
\square\delta\varphi= & [1+3(\frac{m_{\varphi}}{M})^{2}\{1-(\frac{m_{\varphi}}{M})^{2}\}]^{-1}\times\nonumber \\
 & [m_{\varphi}^{2}-(\frac{m_{\varphi}}{M})^{2}\kappa T]\delta\varphi-\kappa(\frac{m_{\varphi}}{M})^{2}\delta T,
\end{align}
where $\delta T$ is a trace of $\delta T_{\mu\nu}^{(\mathrm{M})}$
and $T$ is a trace of $\left\langle T_{\mu\nu}^{(\mathrm{M})}\right\rangle $.

For comparison with the Weyl symmetry breaking in case $f(R)=R$, we recall the result of \citep{21}: 
\begin{equation}
\delta \mathcal{G}_{\mu\nu}=\kappa\delta T_{\mu\nu}^{(\mathrm{M})}+\kappa(\nabla_{\mu}\nabla_{\nu}-g_{\mu\nu}\square)\delta\varphi,
\end{equation}
\begin{equation}
\square\delta\varphi=m_{\varphi}^{2}\delta\varphi.
\end{equation}
These equations correspond to model B in our previous paper, wherein
we adopted the initial value of $\delta\varphi$ as $\delta\varphi_{\mathrm{ini}}=\sqrt{2}A_{s}$
and $(\delta\varphi)_{\mathrm{ini}}^{\cdot}=0$, where $A_{s}$ is
the initial amplitude of the scalar fluctuation. With the limit $m_{\varphi}\rightarrow\infty$,
(62) and (63) approach the results of GR. However, we expect that
with such initial values, we cannot find any reasonable limit by which
we can obtain the results of GR because of the term $(m_{\varphi}/M)^{2}$.
Hence, for the initial values of the current theory (60) and (61),
we adopt the following: 
\begin{equation}
\delta\varphi_{\mathrm{ini}}=0,
\end{equation}
\begin{equation}
(\delta\varphi)_{\mathrm{ini}}^{\cdot}=0.
\end{equation}
Note that in our previous study of model B, all the observational
results would be equivalent to GR with this choice of initial values,
as there can be no evolution of the Weyl field. However, since we
now have an additional source term $\delta R$ in the equation of
motion for the Weyl field, we expect that there must be some evolution
of the Weyl field, even with this simplest initial value choice. In
addition, one can clearly see that one can obtain the results of GR
in the limit $m_{\delta\varphi}\rightarrow0$, contrary to our previous
study.

From now on, we establish how one can observationally verify the Weyl symmetry breaking in the Starobinsky inflation and check the expectations
we made above. For numerical computations, we have used CLASS, which
computes perturbational quantities in the cosmological background and
CMB multipoles at high accuracy \citep{class}. \footnote{We also used
the Fortran code CAMB for reference: \citep{26}.} For numerical values
in the code, we have used Planck 2018 best-fit values \citep{27}:
$H_{0}=67.32$ $\Omega_{b}h^{2}=0.022383$, $\Omega_{c}h^{2}=0.12011$,
$\tau=0.0543$, $\ln(10^{10}A_{s})=3.0448$, $n_{s}=0.96605$ and
assumed the homogeneous and isotropic universe with a cosmological
constant and a flat spatial curvature, i.e., $K=0$.

First, we plot the CMB TT spectrum by varying the Weyl field mass
and the scalar-to-tensor ratio $r$ to understand how the change manifests
in CMB observables.

\begin{figure}
\makebox[\linewidth]{
\begin{centering}
\includegraphics[scale=0.4]{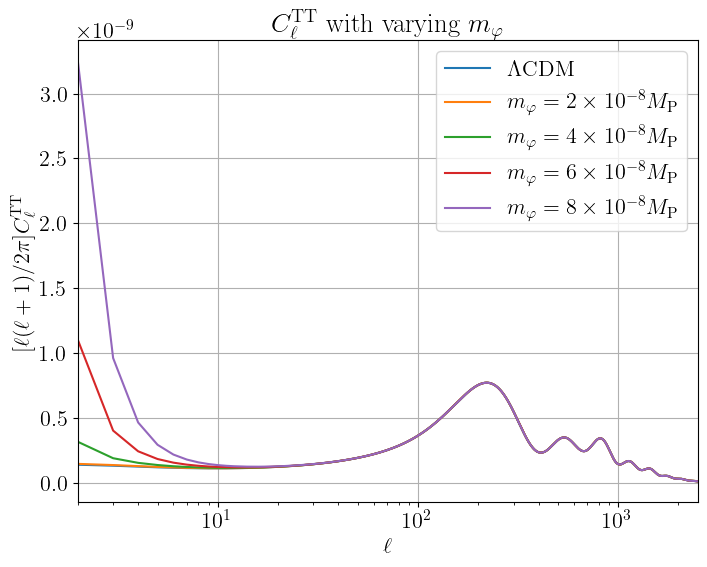}\includegraphics[scale=0.4]{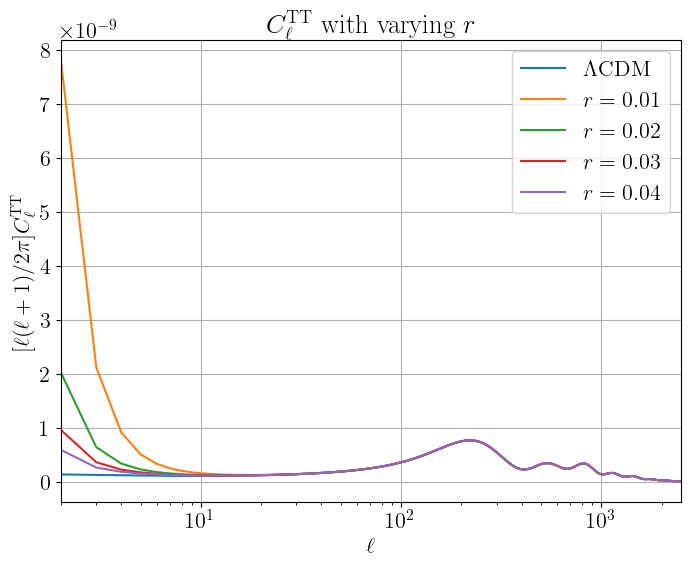}
\par\end{centering}
}
\caption{CMB temperature power spectra $\ell(\ell+1)C_{\ell}/2\pi$ in a unit
of $\mu\mathrm{K}^{2}$, with varying values of the Weyl field mass
$m_{\varphi}$ and the scalar-to-tensor ratio $r$. As $m_{\varphi}$
increases $r$ decreases, the power spectra at the low-$\ell$ region
increase dramatically. The spectra at high-$\ell$ region($\ell\gtrsim10^{2}$)
shows much fewer changes.}
\end{figure}

As Figure 1 shows, the effect of symmetry breaking is most dominant
in the low-$\ell$ region of the multipoles. As the Weyl field mass
$m_{\varphi}$ increases, the values at the low-$\ell$ region grow up
dramatically, whereas they decrease as $r$ gets bigger values. For
the detailed analysis, we investigate how the perturbational values
and multipoles evolved in the universe's early era. To analyze how
this feature is achieved, we adopt the conformal Newtonian gauge:
\begin{equation}
g_{\mu\nu}=-a^{2}(\eta)[\{1+2\Psi(\eta,\mathbf{x})\}\mathrm{d}\eta^{2}+\{1-2\Phi(\eta,\mathbf{x})\}\delta_{ij}\mathrm{d}x^{i}\mathrm{d}x^{j}],
\end{equation}
where $a$ is the scale factor of the universe, and $\mathrm{d}\eta\equiv\mathrm{d}t/a$
is conformal time. Also, we make a normal mode decomposition. First,
let us denote $Q(\mathbf{k},\mathbf{x})$ as an eigenmode of the generalized
Helmholtz equation 
\begin{equation}
\nabla^{2}Q(\mathbf{k},\mathbf{x})=k^{2}Q(\mathbf{k},\mathbf{x}),
\end{equation}
where $\nabla^{2}\equiv a^{2}\delta_{ij}\nabla^{i}\nabla^{i}$ is
a Laplacian and $k^{2}\equiv\left|\mathbf{k}\right|^{2}$. With this
function, we may decompose an arbitrary quantity $X$ as follows:
\begin{equation}
X(\eta,\mathbf{x})=\int\frac{\mathrm{d}^{3}\mathbf{k}}{(2\pi)^{3}}X_{k}(\eta,\mathbf{k})Q(\mathbf{k},\mathbf{x}),
\end{equation}
where the subscript $k$ denotes that it is a decomposed form. From
now on, we assume all the quantities are decomposed and suppress the
subscript unless needed. From the Einstein equation (60), the
anisotropic stress part is given as follows: 
\begin{align}
k^{2}(\Phi-\Psi)= & 12\pi G(\bar{\rho}+\bar{p})\pi_{k}+\nonumber \\
 & +[1-(\frac{m_{\varphi}}{M})^{2}]k^{2}\delta\varphi,
\end{align}
where $\bar{\rho}$ and $\bar{p}$ are the average energy density
and pressure and $(\rho+p)\pi_{k}\equiv-(\hat{k}^{i}\hat{k}^{j}-\delta^{ij}/3)\Pi_{ij}$
where $\hat{k}_{i}$ is the unit vector of $\mathbf{k}$ and $\Pi_{ij}$
is the anisotropic part of the energy-momentum tensor. To connect
metric fluctuations with the CMB temperature anisotropy, let us expand
the photon energy density perturbation $\delta_{\gamma}\equiv1-\rho_{\gamma}/\bar{\rho}_{\gamma}$
with Legendre polynomials $\mathcal{P}_{l}(\mu)$, where $\rho_{\gamma}$
and $\bar{\rho}_{\gamma}$ are the photon energy density and its average.
\begin{equation}
\Theta_{l}=(-i)^{-l}\int_{-1}^{-1}\frac{\mathrm{d}\mu}{2}\mathcal{P}_{l}(\mu)\delta_{\gamma}(\mu),
\end{equation}
where $\mu=\mathbf{k}\cdot\hat{\mathbf{z}}$ is a projection of the
vector $\mathbf{k}$ to the $z$-axis. The CMB TT power spectra $C_{l}$
is given by 
\begin{equation}
C_{l}=4\pi\int\frac{\mathrm{d}k}{k}\,\left|\Theta_{l}(\eta_{0},k)\right|^{2}\Delta_{\mathcal{R}}^{2}(k),
\end{equation}
where $\eta_{0}$ is a current conformal time, and $\Delta_{\mathcal{R}}(k)$
denotes dimensionless matter power spectra. Meanwhile, it is well
known that $\Theta_{l}$ can be separated into a few parts \citep{28}.
One of them is called ISW part, written as 
\begin{equation}
\Theta_{l}^{(\mathrm{ISW})}\equiv\int_{\eta_{*}}^{\eta_{0}}\mathrm{d}\eta\,[\Phi^{\prime}(k,\eta)+\Psi^{\prime}(k,\eta)]j_{l}(k(\eta_{0}-\eta)),
\end{equation}
where $\eta_{*}$ is a conformal time at decoupling and $j_{l}$
is a spherical Bessel function. The critical point of the ISW term
is that it contributes to CMB power spectra when the gravitational
potential, i.e., $\Phi$ or $\Psi$, varies through time. In GR, it
is known that this phenomenon happens only in the radiation-dominated
era and the dark-energy-dominated era. From (69), we may write 
\begin{align}
\Theta_{l}^{(\mathrm{ISW,Weyl})}= & \Theta_{l}^{(\mathrm{ISW,GR})}\nonumber \\
 & -[(\frac{m_{\varphi}}{M})^{2}-1]\int_{\eta_{*}}^{\eta_{0}}\mathrm{d}\eta\,\delta\varphi^{\prime}j_{l}(k(\eta_{0}-\eta)),
\end{align}
where $\Theta_{l}^{(\mathrm{ISW,Weyl})}$ denotes the ISW effect in
our current model and $\Theta_{l}^{(\mathrm{ISW,GR})}$ denotes one
in GR. From (66), we expect that effects in CMB TT multipoles come
from how the time derivative of the perturbed Weyl field evolves through
time.
\begin{figure}
\makebox[\linewidth]{
\centering{}\includegraphics[scale=0.4]{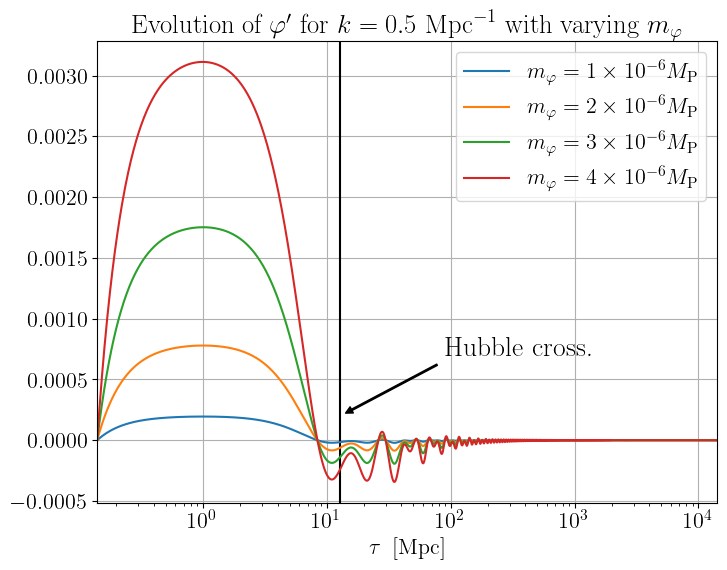}\includegraphics[scale=0.4]{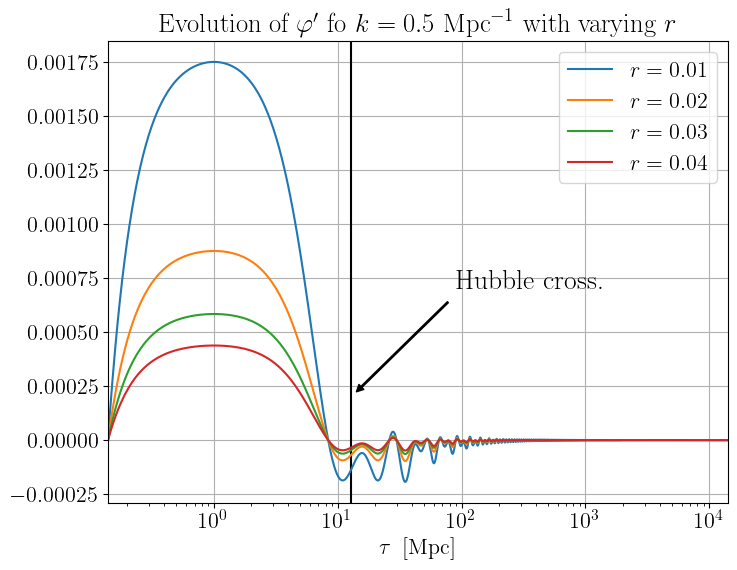}
}
\caption{Evolution of the time derivative of the perturbed Weyl field $\delta\varphi$
through time for $k=0.5$, with varying values of the Weyl field mass
$m_{\varphi}$ and the scalar-to-tensor ratio $r$. Time is in units
of Megaparsec. The black vertical line denotes when the field
crosses the Hubble horizon, and the field drastically shrinks after crossing
the Hubble horizon.}
\end{figure}

Figure 2 shows how $\delta\varphi^{\prime}$ varies with the values
of $m_{\varphi}$ and $r$. It evident from the evolution equation
(62) that the field has more oscillation when the mass of the field
gets heavier and the tensor amplitude becomes smaller. In addition,
the oscillation directly shrinks after the field crosses the Hubble
horizon. That is, this feature is more notable in the early era of
the universe. This is not surprising because the source of the oscillation
largely comes from the photon fluctuation, which tends to hold non-zero
values inside the Hubble horizon but becomes damped out beyond the
Hubble crossing. In the late-time, on the other hand, the effects
from the oscillation would cancel out themselves because they tend
to be negative for the equal possibility, just as the values are likely
to be positive. 
\begin{figure}
\makebox[\linewidth]{
\centering{}\includegraphics[scale=0.4]{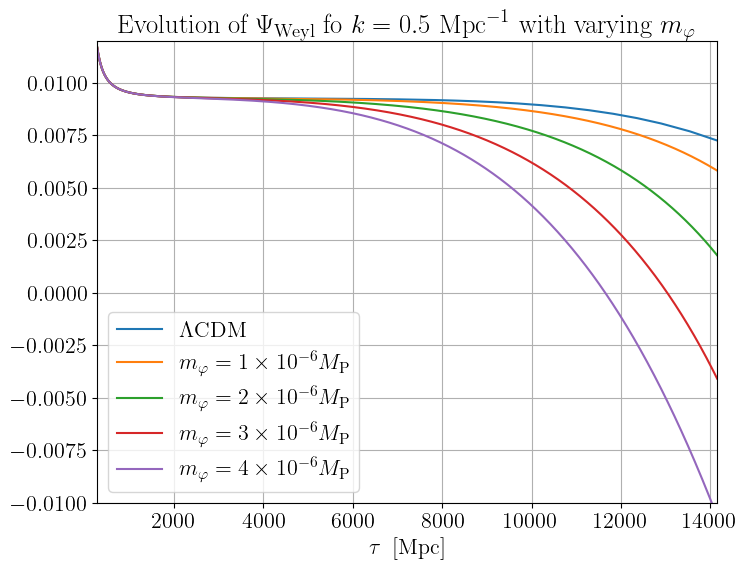}\includegraphics[scale=0.4]{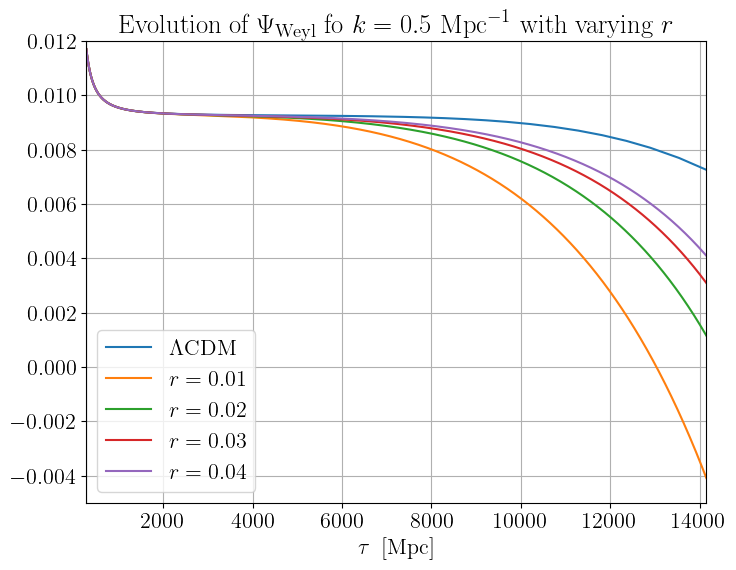}
}
\caption{Evolution of the Weyl potential $\Psi_{\mathrm{Weyl}}$ through time
for $k=0.5$, with varying values of the Weyl field mass $m_{\varphi}$
and the scalar-to-tensor ratio $r$. Time is in units of Megaparsec.}
\end{figure}
\begin{figure}
\makebox[\linewidth]{
\begin{centering}
\includegraphics[scale=0.4]{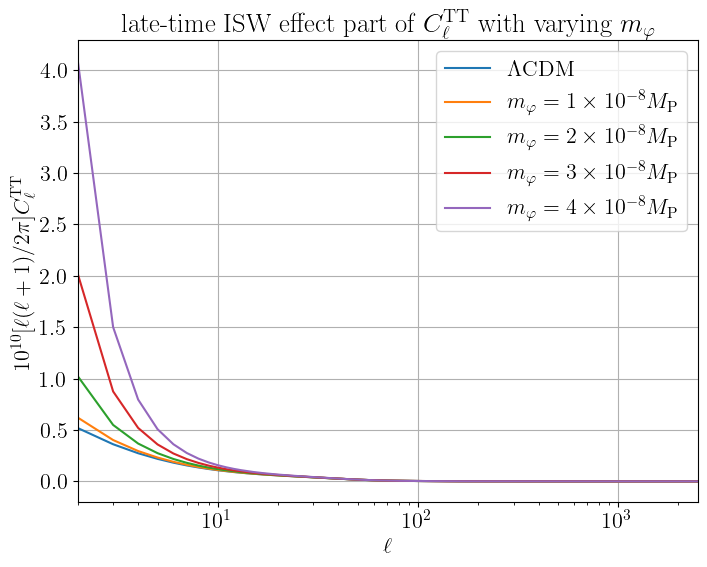}\includegraphics[scale=0.4]{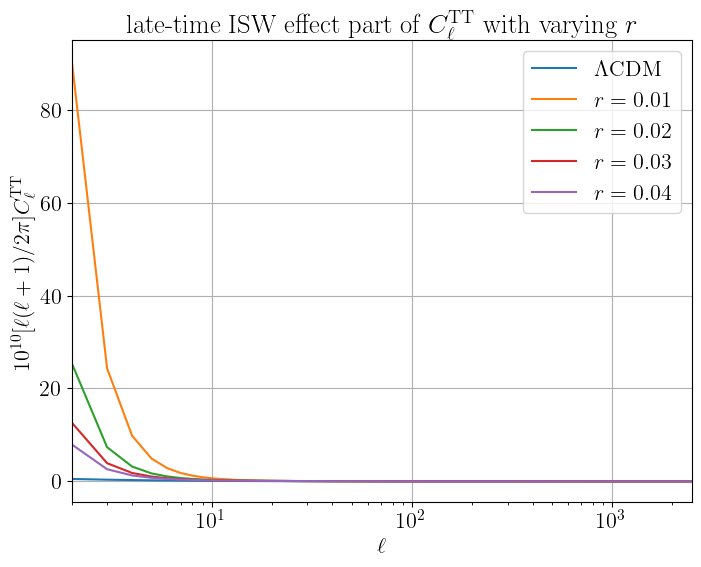}
\par\end{centering}
}
\caption{Integrated Sachs-Wolfe(ISW) effect in CMB temperature power spectra
in a unit of $\mu\mathrm{K}^{2}$, with varying values of the Weyl field
mass $m_{\varphi}$ and the scalar-to-tensor ratio $r$. Note that we have multiplied a factor of $10^{10}$ by $C_{\ell}$ for convenience.}
\end{figure}

 In Figure 3, we also plot how the Weyl potential evolves through
time, defined by
\begin{equation}
\Psi_{\mathrm{Weyl}}\equiv\Phi^{\prime}(k,\eta)+\Psi^{\prime}(k,\eta),
\end{equation}
to show how the effect from the Weyl field manifests in the metric
fluctuation. It is easy to notice that there appear additional oscillative
features in the Weyl potential in the late era. The reason why the oscillation
is more intensive in the late era is that the effect of the Weyl field is integrated through time, which is clear when one observes
the equation (73). Hence, we conclude that the late-time ISW effect
is dominant in CMB temperature anisotropy multipoles, where we also
demonstrate the feature graphically in Figure 4.

 In Figure 5, we plot the CMB polarization EE spectrum and its cross-spectrum
with the temperature TE spectrum. The figure shows that changes
with varying $m_{\varphi}$ in EE and TE spectra are relatively negligible
compared with the TT spectra. In Figure 6, we also plot the CMB polarization
BB spectra. Basically, they show no other extraordinary feature
and thus have less importance than the TT spectra. There are some small
deviations in the plot varying $r$, but they directly originate
from the change of the GW amplitude rather than the effect of the
symmetry breaking.

\begin{figure}
\makebox[\linewidth]{
\begin{centering}
\includegraphics[scale=0.4]{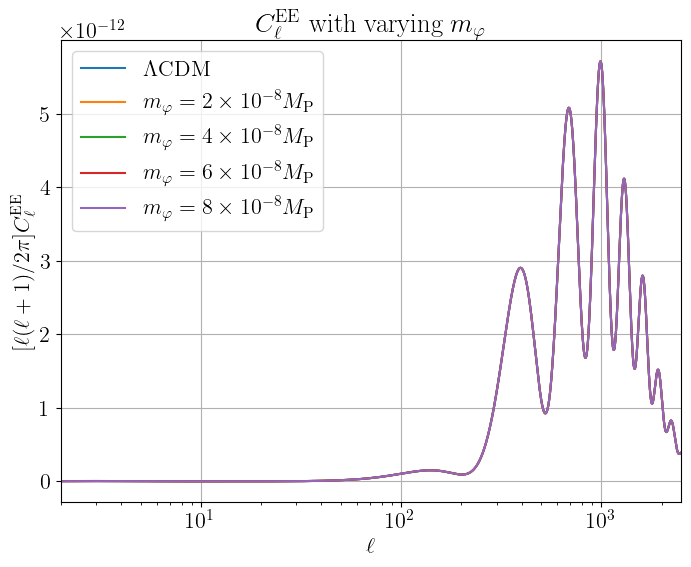}\includegraphics[scale=0.4]{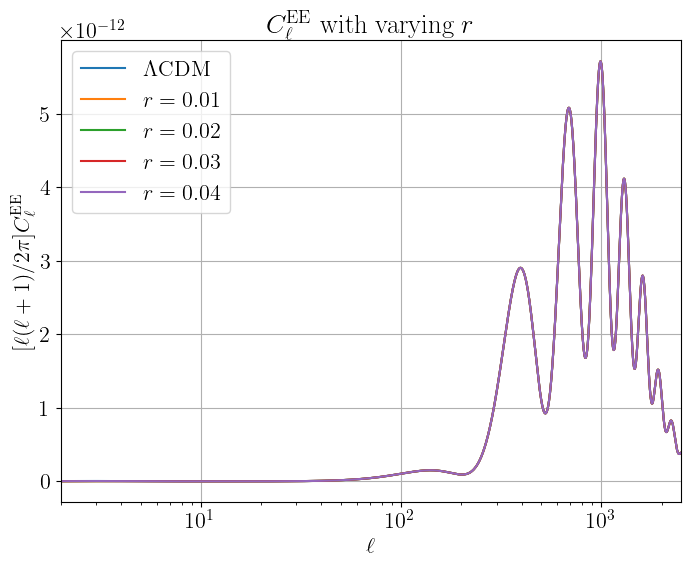} 
\par\end{centering}
}
\makebox[\linewidth]{
\begin{centering}
\includegraphics[scale=0.4]{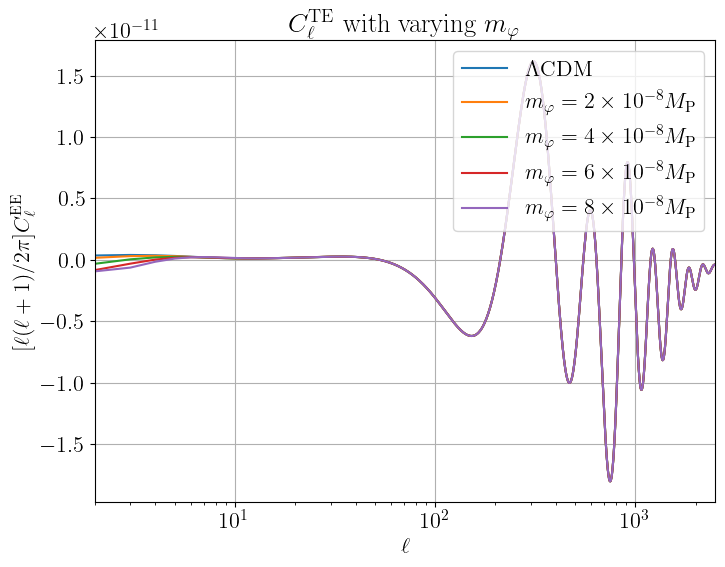}\includegraphics[scale=0.4]{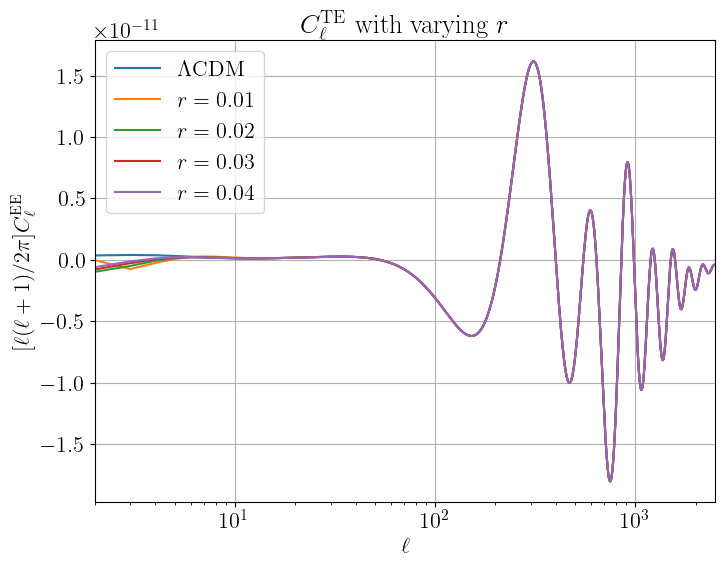}
\par\end{centering}
}
\caption{CMB EE and TE power spectra in a unit of $\mu\mathrm{K}^{2}$, with
varying values of the Weyl field mass $m_{\varphi}$ and the scalar-to-tensor
ratio $r$.}
\end{figure}
\begin{figure}
\makebox[\linewidth]{
\begin{centering}
\includegraphics[scale=0.4]{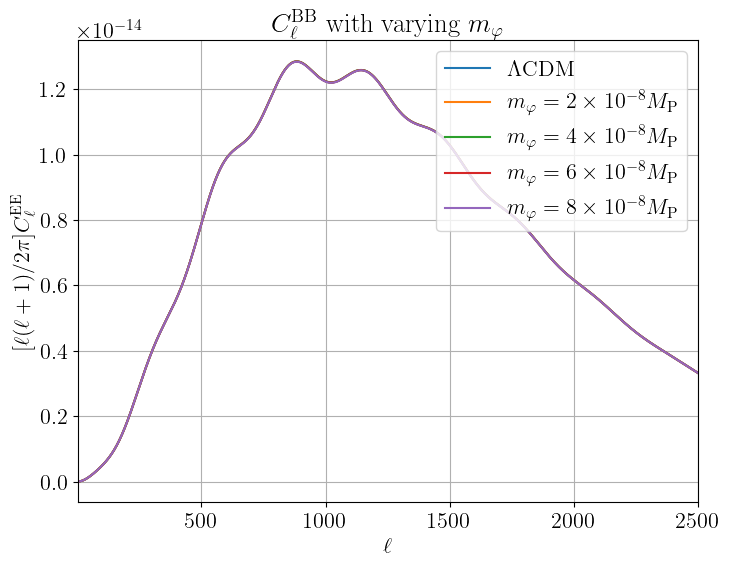}\includegraphics[scale=0.4]{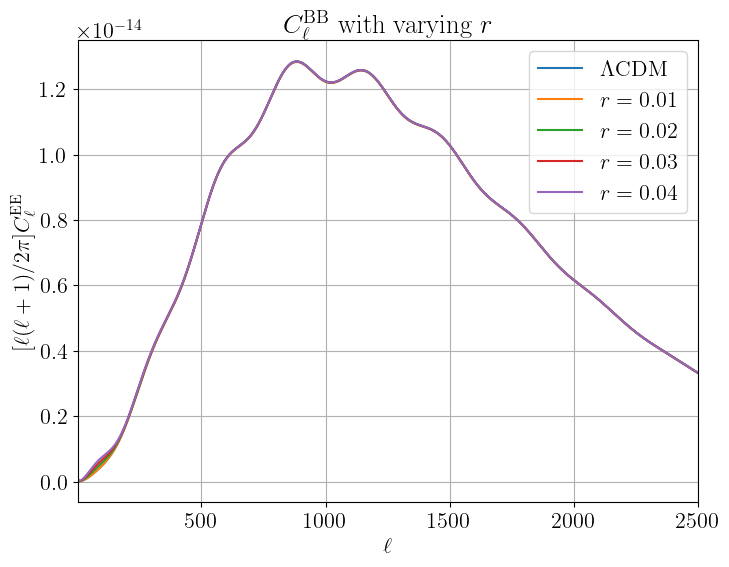}
\par\end{centering}
}
\caption{CMB BB power spectra in a unit of $\mu\mathrm{K}^{2}$, with varying
values of the Weyl field mass $m_{\varphi}$ and the scalar-to-tensor
ratio $r$.}
\end{figure}

To investigate how these modifications affect the evaluation of cosmological
variables, we convey a Bayesian Monte-Carlo simulation with the public
python code Cobaya \citep{JesusTorrado2021,Torrado2019}. We have
run a multiple of chains and checked a convergence with Gelmann-Rubin
statistics $R-1<0.01$ and used data sets from Planck collaboration,
including lensing data \citep{Collaboration2020,Collaboration2020a},
and Bicep/Keck collaboration \citep{Collaboration2021}. Since the
changes in the ISW effect may give us a different value of the current
Hubble parameter when comparing models with data, one may expect that
this might be a novel solution to Hubble tension, one of the most
notorious problems in cosmology today. However, we see almost no difference
in the values of the Hubble parameter in Table 1 and Figure 7. Rather,
as predicted before, our model in this section tends to allow a much
bigger value for the tensor-to-scalar ratio at $k=0.05 \mathrm{Mpc}^{-1}$ because smaller values
for $r_{0.05}$ would contribute to the more considerable ISW effect, which brings
the computed CMB anisotropy values deviate from the observational
data. In Figure 8, we further show what differences appear when using
two different data sets. The addition of the B-mode data from the
Bicep/Keck data gives a stronger constraint to the upper bound of
the tensor-to-scalar ratio. However, there are fewer effects on
the most probable value, the lower bound of $r_{0.05}$, and the
overall distribution for the Weyl field mass.

Moreover, from the table
1, the result does not allow the absence of the primordial GW and
even has a nonzero minimum bound for them under the 68\% confidence
region. Of course, we have presumed that the inflationary period exists, so  it is no wonder that the primordial GW should naturally be accompanied. However, the fact that one
has a lower bound value more significant than zero, even negligible, and has a much more profound meaning. This allows us to refute the model or to allow more preference
by further precise observations. By differing from the original
Starobinsky model, which would still allow the possibility of retaining the
model, even though much smaller constraints for the tensor-to-scalar
ratio releases out there, our modification with Weyl symmetry breaking
presents a stricter standard to verify our model. In this respect,
the primordial Weyl gauge symmetry breaking has a significant chance
of being verified by future experiments or observations.

\section{Conclusion}

In this article, we have studied how Weyl gauge symmetry breaking
can arise in $f(R)$ gravity and how this affects the values of the
cosmological variables and the CMB observables. We assumed the connection
satisfies a new kind of geometrical gauge symmetry, and from that,
we have derived the corresponding version of $f(R)$ gravity, satisfying
that symmetry. Next, we invoked the symmetry breaking at the Planck
scale of the universe with the Higgs-type potential for the Weyl scalar
field. We considered the perturbed field $\delta\varphi$ after the
symmetry breaking up to first order in the gravity side of the action
and found an additional non-minimal coupling with the perturbed Weyl
field only valid at the perturbative level.

As an important conclusion,
we have found two significant changes in cosmological observables.
The first one is the value shift of the Planck constant and the cosmological
constant, which may be a new genuine example of the cosmological backreaction
effect. Furthermore, we have discovered that the amplitude of the
primordial GW affects the CMB scalar anisotropy observables, enabling us to verify the model more easily with observational data. We
next studied how Starobinsky inflation works in our model as a case
study. We found no change in the quantum fluctuation of the metric,
and hence the matter power spectra stay the same. However, it appears
that the perturbed Weyl field can affect the CMB observables, especially
the ISW effect in the late era of the universe. We could see that
the CMB TT power spectra increase overall in the low-$\ell$ region.
Although we have not found any clues to resolve cosmological puzzles
such as the Hubble tension, we believe that a much deeper investigation
including other types of $f(R)$ theories might give us some remedies,
or at least some alleviations.

We now compare our results with studies on CMB anisotropy in other
theories. First, we discuss the CMB anisotropy in the original version
of the Brans-Dicke theory, which does not include the potential for the
Brans-Dicke field, There are many studies out regarding Brans-Dicke-type
theories with potential, for example, see: \citep{31,32}. This research
includes many interesting new phenomena, such as an additional ISW
effect like our paper. Nevertheless, this model also requires background
modification, which clearly differs from ours. Especially many
of them are motivated to resolve the dark energy problem. Therefore
it is natural to assume that there will be a difference in the ISW
effect in the late era. In fact, some of the dark energy models in
Brans-Dicke-type theories are suggested to cure the Hubble tension in
the first place. However, in our model, the non-minimal coupling appears
only at the perturbative level; there is no reason to suspect the
change in the ISW part of the CMB spectra at face value. The modified
temporal evolution of the gravitational potential comes purely from
the perturbation theory.

Next, let us briefly review our previous research \citep{21} for
comparison. We have studied how primordial symmetry could be broken
in two different situations. The first one, namely model A, originated
from Zee's broken-symmetric theory of gravity \citep{22}, a Brans-Dicke-type
theory with a Higgs-type potential to invoke primordial symmetry braking,
and the second one, model B, is a modification of the first one by
adopting Palatini formalism to the action of model A. With this method,
we could establish new gravity models with an additional geometrical
gauge symmetry called Weyl geometry. This corresponds to when we set
$f(R)=R$ in equation (15). Contrary to model B, we have not assumed Palatini formalism in our current research and directly adopted
the Weyl gauge symmetry. Then $f(R)$ models satisfying this symmetry
are found. One could apply Palatini formalism also to the action in
the Weyl frame, that is, an action with respect to $g_{\mu\nu}$,
wherein the Weyl field appears as if it has a non-minimal coupling
with the scalar curvature. However, it would bring a different result
that would not coincide with Weyl geometry due to higher-order curvature.
For instance, there is no coupling with the Weyl field for the quadratic
term, i.e., the $R^{2}$ term in our action (15), even in the Weyl
frame. In this case, we would have to study Palatini formalism in
$f(\varphi,R)$ gravity, where there would be no such simple geometrical
symmetry as (4). Model A has the same feature as model B. That
is, the non-minimal coupling is only valid at the perturbative scale.
However, the effect of symmetry was relevant only in the early era
of the universe. In contrast, the evolution of the Weyl field constitutes
the changes in the metric fluctuation in the late era.

Moreover, one must note that we have adopted different initial values
of the perturbed Weyl field and its time derivative of the current
model, i.e., $\delta\varphi_{\mathrm{ini}}=0$ and $(\delta\varphi)_{\mathrm{ini}}^{\cdot}=0$,
whereas, in our previous study, we adopted $\delta\varphi_{\mathrm{ini}}=\sqrt{2}A_{s}$
and $(\delta\varphi)_{\mathrm{ini}}^{\cdot}=0$ to give initial evolution
for the field. In the model in this article, we did not need to adopt
such values since there exists an external source from the ordinary
matters due to the non-minimal coupling from higher curvature terms.
Furthermore, these values would not give us meaningful results for
the CMB observable because of the term $(m_{\varphi}/M)^{2}$ in the
perturbed Einstein equation (53). For the perturbed Weyl field to
have a small enough effect with such nonzero initial conditions we
must impose a significant immense value presumably above Planck
mass, as the more massive the field, the faster it will decrease.  Nevertheless, (60) would then prevent the results from converging
at the limit $m_{\varphi}\rightarrow\infty$. Moreover, the initial
conditions in our model were chosen to avoid this issue; hence one
might think that this is too intentional because in \citep{21}, we
found these non-minimal initial conditions with an analogy to initial
values of the primordial gravitational waves. We suspect this kind
of the problem could be alleviated by adopting a coupling constant for
the Weyl scalar field like $\omega$ in Brans-Dicke theory and leave
this as a future study topic.

Last but not least, we compare our study with the research on
standard $f(R)$ gravity out there. Since many models in $f(R)$ gravity
also require modification of the background equations, we think that
comparing our model with all types of $f(R)$ gravity models would
not be so meaningful. In addition, we focused only on the Starobinsky
model in our paper. So it would be enough only to compare with the
original Starobinsky inflation model. One interesting feature is that
even though we have adopted a new scalar field, there is no change
in the matter power spectra originating from the inflation. However,
we might be able to verify our model by observing the primordial gravitational
waves and the smoking gun for inflation, although it would be nearly impossible
if the perturbed Weyl field has a very light mass. Also, while there are
many studies on Palatini formalism in $f(R)$ gravity and higher-derivative
theories of gravity \citep{33,34,35,36}, there were not so many studies
on Palatini formalism in $f(\varphi,R)$ gravity, primarily focused
on the geometrical symmetry. Hence, it would be interesting to study
an application of Palatini formalism in $f(\varphi,R)$ gravity to
find their new possible symmetry and its breaking phenomena.

Before closing our manuscript, let us reflect on the research we have
conducted. First of all, our study exploits some extraordinary assumptions,
which might be harmful in some cases. We have especially assumed that the coupling constant $\alpha$ is small, so one can directly connect $\varphi$ with $\phi$. Also, we have used an
approximated equation of motion (21) when we find an equivalent action
with a scalar field. This strategy was to reduce the higher-order
derivatives in our theory. However, in principle, one must abandon
such an assumption to construct a completely exact form of the theory.
Also, we had to adopt Weyl geometry in the first place and did not
use Palatini formalism, which gave us the original motivation to extend
the symmetry of GR \citep{15}. Of course, there is no fundamental
difference or advantage to using Palatini formalism. However, at least
it would be worth applying it to action (8) to investigate
whether there would be some new class of geometrical gauge symmetry.

The physics at the energy scale we have tried to deal with is still totally
unknown to us. And, what is worse, although the higher-order correction of the action
we considered could arise from the high-energy quantum correction,
this would turn out to be, of course, not true. Furthermore, it is
even needless to say that the inflationary scenario itself is still
not fully confirmed, although many cosmologists today believe it would
be true. Nevertheless, we might be able to think of our model as one
of the practical descriptions of gravity under the assumption that
there existed an inflationary period during the evolution of the universe. Only a more profound study will tell us
whether our idea might turn out to be valid.

\section*{Acknowledgment}
The authors thank Jubin Park and Chaemin Yoon for their valuable comments.
Also, we appreciate Heamin Ko and Eunseok Hwang for helping us use the OMEG computing server. Finally, we would like to show our
gratitude to Antony Lewis for help when using the code CAMB and
Cobaya. This research was supported by the Basic Science Research
Program through the National Research Foundation of Korea, funded
by the Ministry of Education, Science, and Technology (Grant No. NRF-2017R1D1A1B06032249,
Grant No. NRF-2021R1A6A1A03043957). 

\begin{figure}
\makebox[\linewidth]{
\begin{centering}
\includegraphics[scale=0.5]{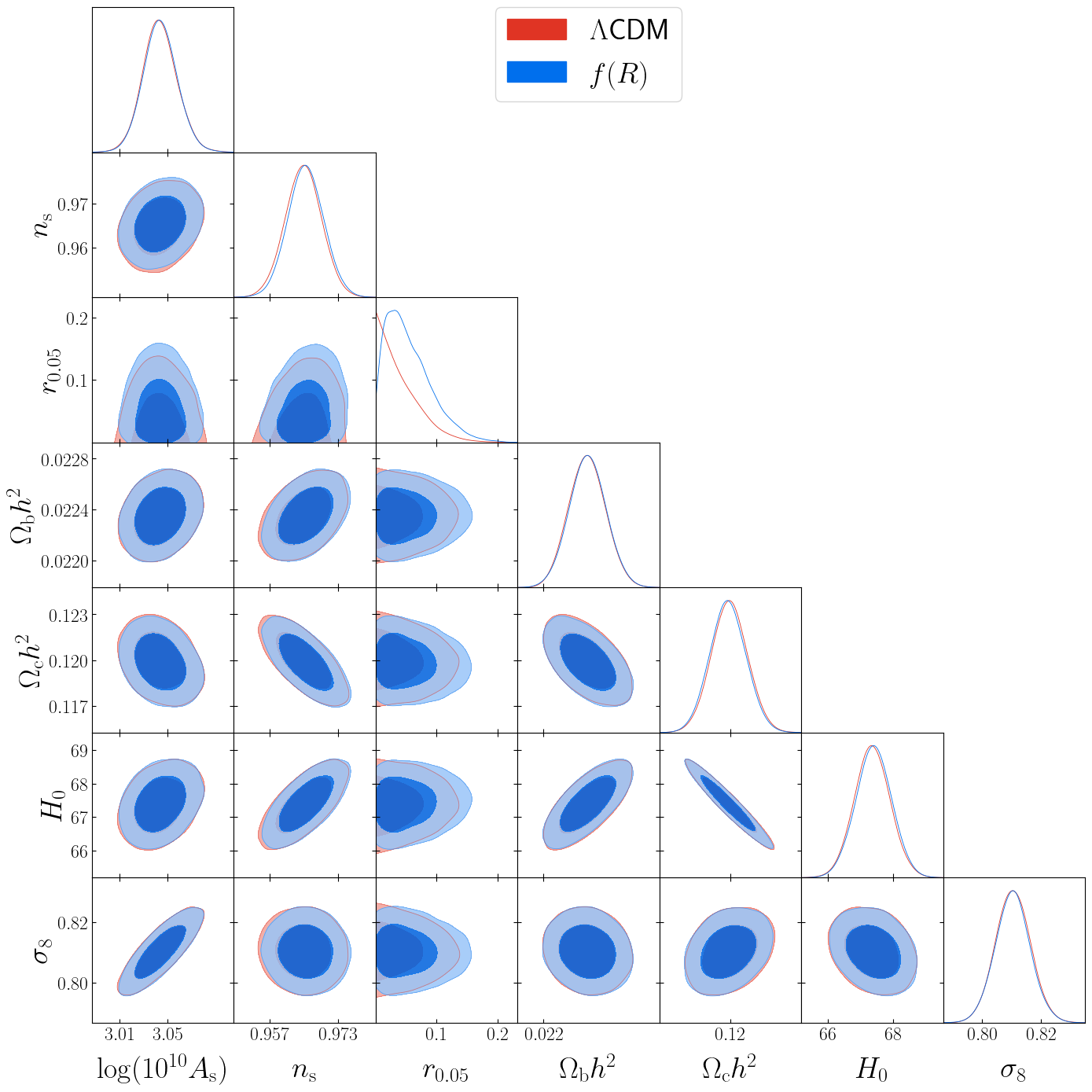}
\par\end{centering}
}
\caption{The 68\% and 95\% contour plots of cosmological parameters estimation
from MCMC simulation using Planck data (with lensing). We see no big
difference between the standard $\Lambda$CDM cosmology(red color)
and our model(blue color) except for the value of the tensor-to-scalar
ratio $r_{0.05}$.}
\end{figure}

\begin{figure}
\makebox[\linewidth]{
\begin{centering}
\includegraphics[scale=0.9]{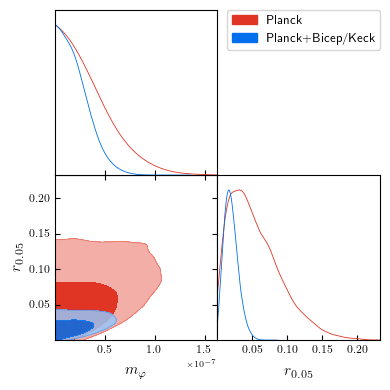}
\par\end{centering}
}
\caption{The contour plots for the Weyl field mass $m_{\varphi}$ in units of the Planck mass(red color)
and the tensor-to-scalar ratio $r_{0.05}$(blue color) using Planck data only
and with Bicep/Keck data. Regions with deeper colors denote 68\% confidence and 95\% confidence regions are depicted with brighter colors. The evaluation with Planck and Bicep/Keck data gives much bigger constraints
on the values of both variables.}
\end{figure}

\begin{table}
\makebox[\linewidth]{
\begin{centering}
\begin{tabular}{|>{\centering}p{2.0cm}|>{\centering}p{2.7cm}|>{\centering}p{2.7cm}|>{\centering}p{2.7cm}|>{\centering}p{2.7cm}|}
\hline 
\multicolumn{5}{|c|}{Planck data}\tabularnewline
\hline 
\hline 
Parameter & $\Lambda$CDM 

(68\% confidence) & $\Lambda$CDM 

(95\% confidence) & $f(R)$

(68\% confidence) & $f(R)$

(95\% confidence)\tabularnewline
\hline 
\boldmath$\log(10^{10} A_\mathrm{s})$ & $3.044^{+0.013}_{-0.015}$ & $3.044^{+0.029}_{-0.028}$ & $3.044\pm 0.014$ & $3.044^{+0.029}_{-0.028}$\tabularnewline
\hline 
\boldmath$n_\mathrm{s}$ & $0.9649\pm 0.0042$ & $0.9649^{+0.0083}_{-0.0084}$ & $0.9655\pm 0.0042$ & $0.9655^{+0.0083}_{-0.0081}$\tabularnewline
\hline 
\boldmath$r_{0.05}$ & $< 0.0505$ & $< 0.107$ & $0.055^{+0.017}_{-0.049}$ & $< 0.127$\tabularnewline
\hline 
\boldmath$\Omega_\mathrm{b} h^2$ & $0.02235\pm 0.00015$ & $0.02235^{+0.00029}_{-0.00028}$ & $0.02235\pm 0.00015$ & $0.02235^{+0.00029}_{-0.00029}$\tabularnewline
\hline 
\boldmath$\Omega_\mathrm{c} h^2$ & $0.1200\pm 0.0012$ & $0.1200^{+0.0024}_{-0.0024}$ & $0.1198\pm 0.0012$ & $0.1198^{+0.0024}_{-0.0024}$\tabularnewline
\hline 
\boldmath$\tau_\mathrm{reio}$ & $0.0541\pm 0.0074$ & $0.054^{+0.015}_{-0.014}$ & $0.0545\pm 0.0074$ & $0.054^{+0.015}_{-0.014}$\tabularnewline
\hline 
$H_0$ & $67.36\pm 0.55$ & $67.4^{+1.1}_{-1.1}$ & $67.41\pm 0.55$ & $67.4^{+1.1}_{-1.1}$\tabularnewline
\hline 
$\sigma_8$ & $0.8105\pm 0.0060$ & $0.810^{+0.012}_{-0.012}$ & $0.8104\pm 0.0060$ & $0.810^{+0.012}_{-0.012}$\tabularnewline
\hline 
\end{tabular} 
\par\end{centering}
}

\makebox[\linewidth]{
\begin{tabular}{c}
\\
\end{tabular}
}

\makebox[\linewidth]{
\begin{centering}
\begin{tabular}{|>{\centering}p{2.0cm}|>{\centering}p{2.7cm}|>{\centering}p{2.7cm}|>{\centering}p{2.7cm}|>{\centering}p{2.7cm}|}
\hline 
\multicolumn{5}{|c|}{Planck+Bicep/Keck data}\tabularnewline
\hline 
\hline 
Parameter & $\Lambda$CDM

(68\% confidence) & $\Lambda$CDM

(95\% confidence) & $f(R)$

(68\% confidence) & $f(R)$

(95\% confidence)\tabularnewline
\hline 
\boldmath$\log(10^{10} A_\mathrm{s})$ & $3.045\pm 0.014$ & $3.045^{+0.029}_{-0.027}$ & $3.045\pm 0.014$ & $3.045^{+0.028}_{-0.027}$\tabularnewline
\hline 
\boldmath$n_\mathrm{s}$ & $0.9643\pm 0.0041$ & $0.9643^{+0.0082}_{-0.0080}$ & $0.9645\pm 0.0041$ & $0.9645^{+0.0082}_{-0.0080}$\tabularnewline
\hline 
\boldmath$r_{0.05} $ & $0.0161^{+0.0061}_{-0.013}$ & $< 0.0348$ & $0.0196^{+0.0076}_{-0.012}$ & $0.020^{+0.020}_{-0.019}$\tabularnewline
\hline 
\boldmath$\Omega_\mathrm{b} h^2$ & $0.02234\pm 0.00015$ & $0.02234^{+0.00029}_{-0.00028}$  & $0.02235\pm 0.00014$ & $0.02235^{+0.00029}_{-0.00028}$\tabularnewline
\hline 
\boldmath$\Omega_\mathrm{c} h^2$ & $0.1201\pm 0.0012$ & $0.1201^{+0.0023}_{-0.0023}$ & $0.1201\pm 0.0012$ & $0.1201^{+0.0023}_{-0.0023}$\tabularnewline
\hline 
\boldmath$\tau_\mathrm{reio}$ & $0.0545^{+0.0067}_{-0.0075}$ & $0.054^{+0.015}_{-0.014}$ & $0.0545\pm 0.0074$ & $0.055^{+0.015}_{-0.014}$\tabularnewline
\hline 
$H_0$ & $67.30\pm 0.53$ & $67.3^{+1.0}_{-1.0}$ & $67.32\pm 0.53 $ & $67.3^{+1.0}_{-1.0}$\tabularnewline
\hline 
$\sigma_8$ & $0.8113\pm 0.0057$ & $0.811^{+0.012}_{-0.011}$ & $0.8111\pm 0.0058$ & $0.811^{+0.012}_{-0.011}$\tabularnewline
\hline 
\end{tabular}
\par\end{centering}
}
\caption{The best-fit values for main cosmological parameters and their 68\% and
95\% confidence bounds, using Planck data only and Planck+Bicep/Keck
data. There are subtle and not significant changes except for the
tensor-to-scalar ratio $r_{0.05}$. As expected, it will likely have
a bigger value when assuming the Starobinsky inflation with the Weyl
symmetry breaking.}

\end{table}

 \bibliographystyle{unsrt}
\bibliography{reference}

\end{document}